%% file: main.tex
\documentclass[10pt,twocolumn,letterpaper]{article}

\usepackage[pagenumbers]{cvpr} 
\usepackage{graphicx}

\input{preamble}

\definecolor{cvprblue}{rgb}{0.21,0.49,0.74}
\usepackage[pagebackref,breaklinks,colorlinks,citecolor=cvprblue]{hyperref}

\title{Anatomically Constrained Implicit Face Models}

\author{Prashanth Chandran\\
	DisneyResearch\textbar Studios \\
	{\tt\small {prashanth.chandran@disneyresearch.com}}
\and
	Gaspard Zoss\\
	DisneyResearch\textbar Studios \\
	{\tt\small {gaspard.zoss@disneyresearch.com}}
}

\input{macros.tex}

\begin{document}
\maketitle
\input{0_abstract}    
\input{1_intro}
\input{2_relatedwork}

\input{3_method}
\input{4_results}
\input{5_conclusion}
{
    \small
    \bibliographystyle{ieeenat_fullname}
    \bibliography{main}
}

\appendix

\input{supplemental}
\end{document}

%% file: preamble.tex
\usepackage[dvipsnames]{xcolor}

%% file: macros.tex
\usepackage{color}

\newcommand{\figref}[1]{Fig.~\ref{#1}}
\newcommand{\tabref}[1]{Table~\ref{#1}}
\newcommand{\eqnref}[1]{Eq.~\ref{#1}}
\newcommand{\secref}[1]{Section~\ref{#1}}

%% file: 0_abstract.tex
\begin{abstract}
Coordinate based implicit neural representations have gained rapid popularity in recent years as they have been successfully used in image, geometry and scene modeling tasks. In this work, we present a novel use case for such implicit representations in the context of learning anatomically constrained face models. Actor specific anatomically constrained face models are the state of the art in both facial performance capture and performance retargeting. Despite their practical success, these anatomical models are slow to evaluate and often require extensive data capture to be built. We propose the anatomical implicit face model; an ensemble of implicit neural networks that jointly learn to model the facial anatomy and the skin surface with high-fidelity, and can readily be used as a drop in replacement to conventional blendshape models. Given an arbitrary set of skin surface meshes of an actor and only a neutral shape with estimated skull and jaw bones, our method can recover a dense anatomical substructure which constrains every point on the facial surface. We demonstrate the usefulness of our approach in several tasks ranging from shape fitting, shape editing, and performance retargeting. 
\end{abstract}

%% file: 1_intro.tex
\section{Introduction}
\label{sec:intro}

Deformable face models are an important tool in the arsenal of visual effects artists dealing with facial animation. As they are ubiquitously used both in high-end production workflows and lightweight consumer applications, building expressive face models for various applications continues to remain an active area of research~\cite{Egger2020}.
Face models today can range from simple linear global shape models~\cite{Blanz1999,Lewis2014,FLAME2017} to highly complex local models that incorporate the underlying facial anatomy through physical simulation~\cite{Yang2022,Animatomy,SoftDECA} or through anatomical constraints~\cite{Wu2016ALM}. 

In this work, we concern ourselves primarily with the high-quality facial animation workflow where actor specific linear blendshape models~\cite{Lewis2014} continue to remain the most commonly used tool for creating facial animations~\cite{FacialRigs2012,Wu2016ALM,Karacast2022}. We propose a new class of actor specific shape models named the \emph{Anatomical Implicit face Model} (AIM) which provides several unique advantages over the existing actor specific face models, and can be used as a drop-in replacement for traditional blendshape models.

An actor specific blendshape model is a collection of 3D shapes of the given actor performing a number of facial expressions, usually created by face scanning~\cite{Beeler:2011} or by an artist. While the user-friendliness of such actor specific blendshape models contributes to their wide adoption, it is a well known limitation that such models often require hundreds of shapes to accurately model complex facial deformation~\cite{Lewis2014}. To address these shortcomings, local blendshape models~\cite{Tena2011,Wu2016ALM,Karacast2022} were proposed. By splitting the face into regions, and allowing the individual regions to deform independently, local shape models are able to capture complex deformations with a limited number of shapes.

While local models address the lack of expressivity in global shape models, state-of-the-art methods in facial performance capture~\cite{Wu2016ALM} and retargeting~\cite{Karacast2022} often incorporate anatomical constraints on the facial surface to plausibly restrict the range of the skin deformations. The anatomical constraints employed by these models~\cite{Wu2016ALM,Karacast2022}  provide a few hidden advantages that end up contributing towards their practical success. For example, in the context of facial performance capture, Wu \etal~\cite{Wu2016ALM} demonstrated that including anatomical constraints derived from the relationship between the facial skin and underlying bones (skull and mandible) helps to separate the rigid and non-rigid components of facial deformation, leading to better face performance capture. In the context of facial performance retargeting, Chandran \etal~\cite{Karacast2022} made use of such an anatomically constrained local face model to restrict a retargeted shape to lie within the space of anatomically plausible shapes of the target actor. 

Despite their practical success, anatomical constraints are often formulated in practice as regularization terms that have to be satisfied as part of complex optimization problems involving several objectives. As a result, fitting these anatomical face models to a target scan or an image for instance, is a computationally intensive procedure taking several minutes per frame on a CPU, or requires hand crafted GPU solvers~\cite{fra-2020}. Furthermore anatomy constraints are enforced only in sparse regions of the face, whereas in reality the facial skin surface is more densely constrained by the underlying anatomy, and simulating this dense interaction between the anatomy and facial skin through physical simulation can be even more computationally intensive~\cite{Sifakis2006,Yang2022}. 

In this paper, we propose the \emph{Anatomical Implicit face Model}; a framework that allows for a holistic representation of both the facial anatomy and the skin surface using simple implicit neural networks and facilitates the learning of a continuous anatomical structure that densely constrains the skin surface. Our model formulation, inspired by the anatomical local model (ALM) of Wu \etal~\cite{Wu2016ALM}, can further disentangle deformation arising from rigid bone motion (jaw motion) and non-rigid deformations created by muscle activations. Our model also addresses the computational bottleneck of the ALM model by explicitly deriving the skin surface from the anatomy, instead of formulating it as a constrained optimization problem. By ensuring that a point on the skin surface is always reconstructed through the underlying anatomy, our method provides several unique features in comparison to existing implicit face models, such as anatomy based face manipulation (see \secref{sec:results}). Before describing the details of our anatomical formulation in \secref{subsec:model}, we discuss related work in \secref{sec:relatedwork}. 

%% file: 2_relatedwork.tex
\vspace{-2mm}
\section{Related Work}
\label{sec:relatedwork}
\paragraph{3D Morphable Models} Facial models used in animation make up for an extremely well studied body of work with the earliest works dating back to the late 1970s~\cite{Ekman1978FacialAC}. We therefore refer to the excellent survey of Egger \etal~\cite{Egger2020} for an in-depth review of the state-of-the-art methods, and provide only a concise summary in this section. Facial blendshapes~\cite{Ekman1978FacialAC,Lewis2014} have been conventionally used as a standard tool by artists to navigate the geometric space of human faces. The seminal 3D linear morphable model proposed by Blanz and Vetter~\cite{Blanz1999} used principal component analysis to describe the variation in facial geometry and texture, which was later extended to multilinear models, jointly modeling identity and expression by Vlasic \etal~\cite{Vlasic2005} and later by Cao \etal~\cite{Cao2014FaceWarehouseA3}. Today a very commonly used morphable face model is the FLAME model~\cite{FLAME2017} which incorporates identity, expression and corrective blendshapes in addition to modeling bone motion with linear blend skinning. Due to its flexible nature, the FLAME model is widely used by face reconstruction algorithms today~\cite{Feng:SIGGRAPH:2021}.  Finally Chai \etal~\cite{REALY} recently created the HIFI3D++ morphable model which is built from a union of scans from several previously proposed models.

In the past few years, numerous face models leveraging the power of deep neural networks to model the nonlinear deformation of the human face have also been proposed. While the initial work in this area by Ranjan \etal~\cite{Ranjan2018} focused on the use of specialized graph convolutional networks to operate on shapes, several later approaches proposed further modifications to the network architecture to improve the accuracy in shape representation \cite{Chen2021,Zhou2020,Bouritsas2019,Gong2019}. To make these deep morphable models intuitive to use, Chandran \etal~\cite{Chandran2020sdfm} subsequently proposed the Semantic Deep Face Model which treats a collection of neural networks like a multilinear model to achieve identity-expression disentanglement. Extensions of such a semantically controllable model to deal with topology changes~\cite{Chandran2022ShapeTransformer} and temporal sequences of geometry~\cite{Chandran2022PerFormer} have also been proposed. Deep neural models that jointly model the facial geometry and appearance with semantic controls have also been proposed~\cite{Ruilong2020}.
\vspace{-4mm}
\paragraph{Implicit Face Models} Owning to the massive success of coordinate based neural networks in representing images~\cite{sitzmann2019siren,ACORN2021}, 3D shapes~\cite{Park2019} and arbitrary scenes~\cite{mildenhall2020nerf}, today's research on parametric face models primarily focuses on implicit representations. Yenamandra \etal~\cite{Yenamandra2021} proposed \emph{i3DMM} as an initial exploration of using coordinate based networks for modeling full head geometries. This was followed by IMFace~\cite{zheng2022imface} which disentangled facial geometry into separate identity and expression embeddings with the help of individual deformation fields. More recently, Neural Parametric Head Models (NPHM)~\cite{giebenhain2023nphm} proposed a method which improves the fidelity of neural implicit representations by jointly training an ensemble of local neural fields centered around anchor points. Implicit neural representations have also successfully been employed in learning an animatable avatar of a human face from only monocular video as demonstrated by IMAvatar~\cite{zheng2022IMavatar} and Point Avatar~\cite{Zheng2023pointavatar}. Wang \etal~\cite{Wang2022} also proposed MoRF, which is a Neural Radiance Field~\cite{mildenhall2020nerf} conditioned on an identity code allowing for photorealistic free viewpoint rendering of the full head in a fixed expression. Recently Buhler \etal~\cite{buhler2023preface} also explored how such multi-identity radiance fields can be fit to sparse images to recover a volumetric head model. Finally coordinate based neural networks have also been successfully employed in creating animatable human body models~\cite{deng2020,NPM2021,Biswas2021,guo2023vid2avatar}.
\vspace{-4mm}
\paragraph{Anatomically Constrained Face Models} The anatomical local model proposed in the context of monocular facial performance capture by Wu \etal~\cite{Wu2016ALM}, first introduced the coupling of the anatomical bone structure to the skin surface and modeled the effect of skin patches sliding over the bone through soft anatomical constraints. This formulation was later adapted by Chandran \etal~\cite{Karacast2022} for facial performance retargeting. Qiu \etal proposed \emph{SCULPTOR}~\cite{Sculptor2022}, a multi-identity joint morphable model of facial anatomy and skin learned from a database of computed tomography (CT) scans. Recently Choi \etal proposed \emph{Animatomy}~\cite{Animatomy}, a muscle fiber based anatomical basis for animator friendly face modeling applications. Lastly we recognize several physically based face models~\cite{Yang2022,SoftDECA,physics2021,Sifakis2006} which inherently have the ability to model anatomy constraints through simulation.

We draw inspiration from the three classes of facial morphable models discussed above and propose the \emph{Anatomical Implicit face Model}: a blendshape based, implicit, anatomically constrained face model targeted towards high-quality actor specific face modeling. Our method can be seen as general extension of local blendshape models~\cite{Karacast2022} to a continuously evaluable implicit function, and represents a set of actor blendshapes through a novel anatomical formulation. Unlike traditional patch-based models, our framework allows us to approximate complex shapes without requiring the user to specify patch layouts and other hyper-parameters. Our solution is based on simple coordinate based MLPs enabling efficient training and inference, and provides computational benefits over previous anatomically formulated face models~\cite{Wu2016ALM}. Finally to the best of our knowledge, our method is the first to explore anatomical constraints inside an implicit facial blendshape model.

%% file: 3_method.tex
\begin{figure}[ht]
	\begin{centering}
		\includegraphics[width=\columnwidth]{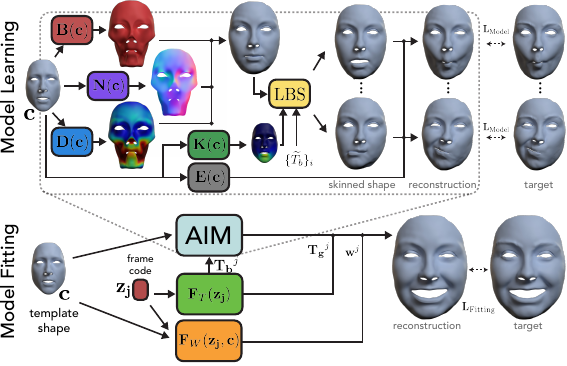}
		\caption{Our approach consists of a model learning stage (\secref{subsec:learning}) and a model fitting stage (\secref{subsec:modelfitting}). In the model learning stage, a set of an actor's blendshapes are memorized by an ensemble of MLPs by our \emph{Anatomical Implicit face Model} (AIM). In the second model fitting stage, the memorized model can be used as power shape prior to fit the actor model to target shapes.}
		\label{fig:overview}
	\end{centering}
	\vspace{-5mm}
\end{figure}
\section{Anatomical Model Formulation}
\label{sec:method}
\label{subsec:model}
The core idea of our approach is to formulate a learning scheme for an implicit neural representation that can reproduce an actor blendshape model while automatically learning the underlying facial anatomy and constraining the skin surface to this learned anatomy. Crucial to our learning scheme is our anatomically constrained face model that geometrically couples the underlying facial anatomy to the enclosing skin surface which we describe next. 

\begin{figure}[ht]
	\begin{centering}
		\includegraphics[width=\columnwidth]{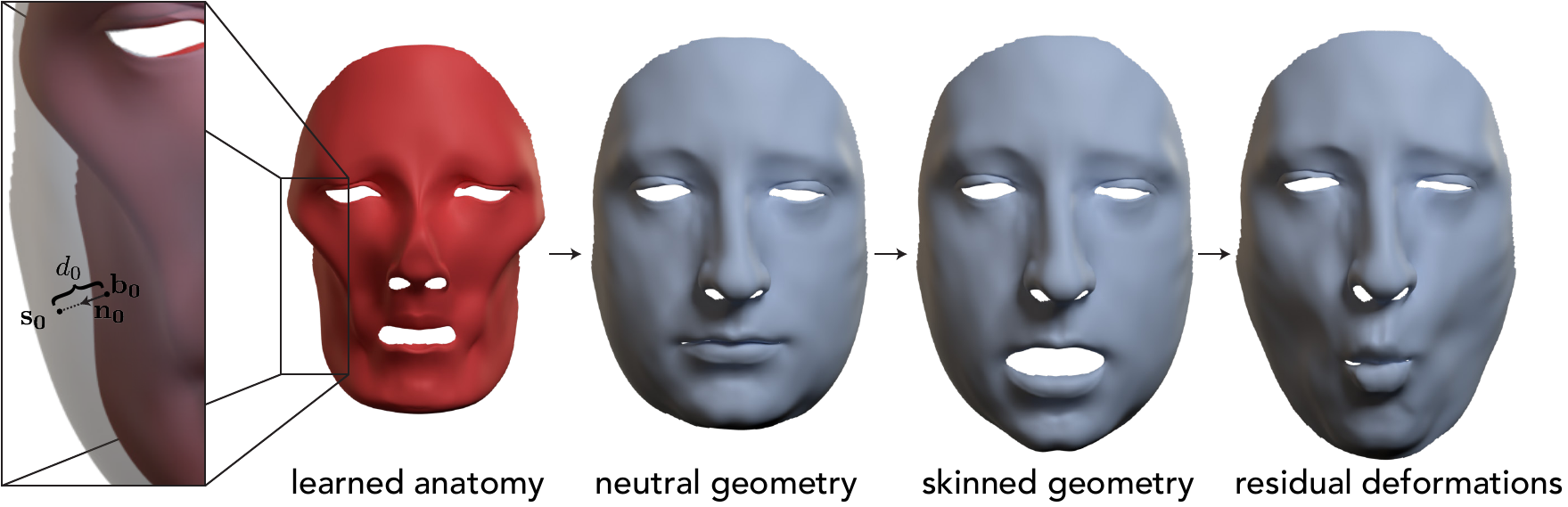}
		\caption{We show the break down of how we anatomically build up the facial skin surface. Starting from a learned anatomy surface (left), and learned anatomic properties like the soft tissue thickness, and anatomic surface normals, we reconstruct the neutral skin geometry. The neutral anatomy is skinned, and non-rigidly deformed with residual displacements to result in the final shape.}
		\label{fig:modelformulation}
	\end{centering}
	\vspace{-5mm}
\end{figure}

We assume that we are given a set of $N$ 3D scans $({S_0, S_1, S_2, .. , S_{N-1}})$ of an actor represented as meshes. Without loss of generality, let $S_0$ be the shape with a neutral expression (or the rest pose). Each shape $S_i$ consists of $V$ vertices, and all shapes share the same vertex connectivity. For simplicity we exclude the index of the vertex in a shape in our notation and present our formulation as operating on surface points $\mathbf{s} \in \mathbf{R}^3$. Let $\mathbf{s_0} \in \mathbf{R}^3$ and $\mathbf{s_i} \in \mathbf{R}^3$ be corresponding points on the skin surface for the neutral expression and expression $i$ respectively. In most previous methods for learning neural face models, a skin surface point $\mathbf{s}$ is learned as a displacement from a base face surface~\cite{Chandran2020sdfm,Chandran2022ShapeTransformer,giebenhain2023nphm} or simply as points lying in an arbitrary 3D space~\cite{zheng2022imface,zheng2022IMavatar,Wang2022}. Contrary to such approaches, we propose to learn the skin surface $\mathbf{s}$ using implicit neural representations that arrives at the facial skin surface through a formulation that combines anatomic constraints, linear blend skinning (LBS), and expression blendshapes into a single framework. 

For our model formulation, we take inspiration from the anatomic constraints first proposed for non-neural face models~\cite{Beeler2014,Wu2016ALM}, particularly that of Wu \etal~\cite{Wu2016ALM}. They establish a link between the skin surface and the anatomic bones by modeling the thickness $d_i \in \mathbf{R}$ of the soft tissue between a bone point $\mathbf{b_i} \in \mathbf{R}$ and the skin surface $\mathbf{s_i}$. These constraints are defined in sparse regions of the face where a skin point can be trusted to have bone underneath. We draw inspiration from their simple formulation and make some important deviations that enable us to jointly learn both the surface of the underlying skin anatomy and the enclosing skin surface for \emph{every} point on the skin through end-to-end learning. Specifically, we arrive at a point on the skin surface as follows
\begin{equation}
	\label{eq:neutral}
	\mathbf{s_0} = \mathbf{b_0} + d_{0} \mathbf{n_0}
\end{equation}
where $\mathbf{s_0}$ is the position of a surface point corresponding to $\mathbf{s}_i$ but on the neutral shape $S_0$, $\mathbf{b_0}$, $d_0$, and $\mathbf{n_0}$ are the bone point, soft tissue thickness and the bone normal at $\mathbf{s_0}$. While \eqnref{eq:neutral} allows us to reconstruct points on the neutral face geometry, to adequately represent skin surfaces under arbitrary facial expressions, we need to account for surface deformation arising from the rigid motion of underlying facial bones (skull and mandible), and the non-rigid skin motion arising from muscle activations, skin sliding, and self collisions. To accommodate these additional degrees of freedom in skin deformation, we incorporate standard linear blend skinning, and expression blendshapes similar to the FLAME model~\cite{FLAME2017}. Therefore given an anatomically reconstructed point on the neutral skin surface $\mathbf{s_0}$, we can now compute the position of the same point in an arbitrary expression $\mathbf{s_i}$ as
	\vspace{-2mm}
\begin{equation}
	\label{eq:skinModel}
	\mathbf{s_{i}} = \text{LBS}(\mathbf{s_0}, T_b, k) + \mathbf{e_i}
\end{equation}
where $\text{LBS}$ refers to the standard linear blend skinning operator that rigidly transforms the anatomically reconstructed neutral surface point $\mathbf{s_0}$ with a transformation $T_b$ and a skinning weight $k$, $\mathbf{e_i} \in \mathbf{R}^3$ is the corrective displacement that is added on top of the skinned result to account for deformations that cannot be explained by skinning alone. A visual overview of our approach to anatomically build up the facial skin surface is shown in \figref{fig:modelformulation}. 

At this point we have established how to arrive at points on the skin surface $\mathbf{s_i}$ for a shape in an arbitrary facial expression $S_i$ by starting from the underlying anatomy $\mathbf{b_i}$. It is important to note that the anatomic constraints as defined by Wu \etal~\cite{Wu2016ALM} can only be computed on regions with an underlying bone, and thus, regions like the cheeks are not anatomically constrained in their approach. An essential feature of our approach that distinguishes it from all previous works is that we enforce anatomic constraints for every point on the skin surface; even in regions where there is no underlying biological bone structure. For this purpose we redefine the anatomy in our work as a rigidly deforming region underneath the skin surface that is not restricted to only the manifold of the skull and mandible bones. Since this structure does not exist in reality and is, therefore, not available for supervised learning, we formulate a learning framework where such rigidly deforming surface can be learnt only from the sparse set of anatomic constraints that can be computed between the skin and the underlying bones. As we will see in \secref{sec:results}, learning this anatomic surface from data leads to several interesting applications in shape manipulation and performance retargeting that were previously challenging to obtain without expensive physical simulation~\cite{Yang2022} or extensive volumetric data capture~\cite{Sculptor2022}. 
\section{Anatomical Implicit Face Model}
At a high level, our method is comprised of two stages: first, a model learning stage (\secref{subsec:learning}) and second, a model fitting stage (\ref{subsec:modelfitting}). In the model learning stage, we bake a collection of expression blendshapes from an actor into an implicit neural network that uses the anatomical model formulation described in \secref{subsec:model}. Our model fitting stage uses this learned \emph{Anatomical Implicit face Model} (AIM) and optimizes for coefficients that deform the model to match test time constraints like 3D shapes, 2D landmarks and so on. The overview of our approach is shown in \figref{fig:overview}.

\begin{figure}
	\begin{centering}
		\includegraphics[width=\columnwidth]{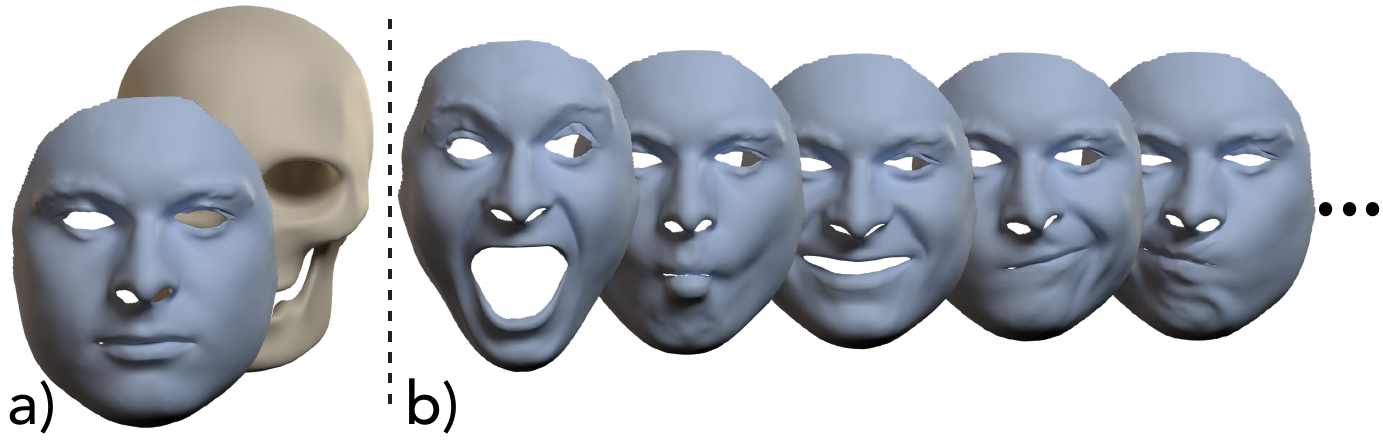}
		\caption{a) We assume we are given the neutral geometry of an actor along with an rough estimate of the skull and jaw bone \cite{Zoss2018}. b) We additionally use a collection of $N$ 3D shapes of the actor performing expressions. Unlike Wu \etal~\cite{Wu2016ALM}, we do not require the tracked anatomy (skull~\cite{Beeler2014}, jaw~\cite{Zoss2019}) for the expression shapes.}
		\label{fig:trainingData}
	\end{centering}
	\vspace{-5mm}
\end{figure}

\subsection{Model Learning}
\label{subsec:learning}
To learn our anatomical implicit face model, we assume we are given a template shape $C$, a registered set of $N$ shapes $({S_0, S_1, S_2, .. , S_{N-1}})$ of a single actor in the same topology of the canonical shape. Additionally we fit a template skull and jaw \emph{only} to the neutral shape using the method of Zoss \etal~\cite{Zoss2018}. The template shape $C$ can either be the neutral shape of the actor or a generic face shape, and the number of shapes provided can be arbitrary. We use a collection of 20 shapes in our work. A visual summary of our training data is shown in \figref{fig:trainingData}. Our objective in the learning stage is to use a coordinate based neural network to memorize the given shapes through the anatomical formulation in  \secref{subsec:model}. Given the high representation power of periodic implicit neural networks~\cite{sitzmann2019siren}, we use the SIREN coordinate network; an MLP with sinusoidal activation functions, as our base architecture. An ablation study on alternate network choices is provided in section \ref{subsec:ablation}.

Given a point $\mathbf{c} \in \mathbf{R}^3$ on the template shape $C$, we use three independent MLPs denoted by $\mathbf{B}$, $\mathbf{D}$, and $\mathbf{N}$ to predict the anatomy point $\mathbf{\widetilde{b}}_0 \in \mathbf{R}^3$, the soft tissue thickness $\widetilde{d}_0 \in \mathbf{R}$, and the anatomy normal $\mathbf{\widetilde{n}}_0 \in \mathbf{R}^3$. These predicted anatomic properties are then used to reconstruct the position of a point on the neutral skin surface $\mathbf{\widetilde{s}}_0$ as 
\begin{eqnarray}
	\mathbf{\widetilde{b}}_0 = \mathbf{B}(\mathbf{c}) \\
	\widetilde{d}_0 = \mathbf{D}(\mathbf{c}) \\
	\mathbf{\widetilde{n}}_0 = \mathbf{N}(\mathbf{c}) \\
	\mathbf{\widetilde{s}}_0 = \mathbf{\widetilde{b}}_0 + \widetilde{d}_{0}\mathbf{\widetilde{n}}_{0}.
\end{eqnarray}
As discussed in \secref{subsec:model}, to further account for the rigid and non-rigid deformations of the skin surface, the anatomically constructed neutral skin point $\mathbf{\widetilde{s}}_0$ has to be skinned and further displaced with residual expression deformations. We therefore employ two additional MLPs $\mathbf{K}$ and $\mathbf{E}$ that predict the skinning weight $\widetilde{k} \in \mathbf{R}$ and the corrective displacements basis $\mathbf{\mathcal{B}_e} \in \mathbf{R}^{(N-1) \times 3}$ respectively. Note here that, as an implementation detail, we predict the expression displacements for all $N-1$ blendshapes (excluding the neutral) at once from $\mathbf{E}$. The corrective expression displacement $\mathbf{\widetilde{e_i}} \in \mathbf{R}^3$ for shape $i$ can be extracted from this output by indexing $\mathbf{\mathcal{B}_e}$ appropriately. 
\begin{eqnarray}
	\widetilde{k} = \mathbf{K}(\mathbf{c}) \\
	\mathbf{\mathcal{B}_e} = \mathbf{E}(\mathbf{c}) \\
	\mathbf{\widetilde{e_i}} = \mathbf{\mathcal{B}_e}[i]\\
	\label{eq:shape}\mathbf{\widetilde{s}}_i = \text{LBS}\left(\mathbf{\widetilde{s}}_0, \widetilde{T_b}, \widetilde{k}\right) + \mathbf{\widetilde{e_i}}
\end{eqnarray}
Here $\widetilde{T_b} \in \mathbf{R}^9$ is a 6-DOF jaw bone transformation optimized along with the training of the MLPs to account for rigid motion of the mandible. Here we parameterize the jaw bone rotation $\widetilde{T_b}$ following the continuous 6D representation~\cite{Zhou2019}.

\subsubsection{Training Objectives}
\label{subsec:trainlearning}
We next describe the training objectives to learn actor expression blendshapes along with the underlying anatomy structure for each skin surface point.\\[5pt]
\textbf{Skin Position Loss} The skin position loss penalizes the difference between the estimated skin point $\mathbf{\widetilde{s}}_i$ and the ground truth skin point $\mathbf{s}_i$.
\begin{equation}
	\mathbf{L}_\text{S} = \lambda_{\text{S}} || \bf{\widetilde{s}}_i - \bf{s}_i ||_{2}^{2}
\end{equation}
We set $\lambda_{\text{S}} = 1.0$ for all our experiments.\\[5pt]
\textbf{Anatomy Regularizer} Since we can roughly estimate the skull and jaw geometry on the neutral shape using the method of Zoss \etal~\cite{Zoss2018}, we compute sparse anatomic constraints~\cite{Wu2016ALM} and loosely regularize the learned anatomic properties to stay close to these estimates only in regions where the constraints can be accurately computed (i.e. skin regions with an underlying bone).
\begin{equation}
	\mathbf{L}_\text{A}\!=\!\lambda_b || \bf{\widetilde{b}}_0 - \bf{b}_0 ||_{2}^{2} + \lambda_d || \widetilde{d}_0 - d_0 ||_{2}^{2} + \lambda_n || \bf{\widetilde{n}}_0 - \bf{n}_i ||_{2}^{2}
\end{equation}
We set $\lambda_{\text{b}} = \lambda_{\text{d}} = \lambda_{\text{n}} = 1.0$ for all our experiments, and observe that this constraint only regularizes 5-10\% of all the vertices generated by the model on average (see Supplemental).\\[5pt]
\textbf{Thickness Regularizer} We regularize the soft tissue thickness $\widetilde{d}$ predicted by the model in unconstrained regions to remain as small unless dictated otherwise by the skin position loss. 
\begin{equation}
	\mathbf{L}_\text{D} = \lambda_{D}^{Reg} || \widetilde{d}_0||_{2}^{2}
\end{equation}
We set $\lambda_{D}^{Reg} = 7.5\mathrm{e}{-4}$ for all our experiments.\\[5pt]
\textbf{Symmetry Regularizer} To exploit the symmetry of the face, we regularize the predictions of the anatomy MLP $\mathbf{B}$ to be symmetric. We achieve this by requiring that reflecting the input points $\mathbf{c}$ along the plane of symmetry provides the same result as reflecting the predicted anatomy points $\mathbf{\widetilde{a}}$.
\begin{equation}
	\mathbf{L}_\text{Sym} = \lambda_{sym} || \bf{B}(\bf{R(c)}) - R(\bf{B}(\bf{c})) ||_{2}^{2}
\end{equation}
where $R$ is an operator that reflects a point along the plane of symmetry. We set $\lambda_{sym} = 1\mathrm{e}{-4}$ for all our experiments. Note that we do not regularize symmetry on the predicted thickness or anatomy normals thereby allowing the model to still be able to represent asymmetric faces.\\[5pt]
\emph{Optional} \textbf{Skinning Weight Regularizer} Finally inspired by~\cite{zheng2022IMavatar}, we use an \emph{optional} loss that encourages the estimated skinning weights $\widetilde{k}$ in regions like the forehead that are guaranteed to not be affected by the rigid deformation of the jaw bone to be zero. 
\begin{equation}
	\mathbf{L}_\text{K} = \lambda_k || \mathbf{K}(\mathbf{c}^*) ||_{2}^{2}
\end{equation}
here $\mathbf{c}^*$ refers to a small region on the canonical shape $C$ which includes the forehead. We set $\lambda_{K} = 1\mathrm{e}{2}$ for all our experiments.

Our final model energy $\mathbf{L}_\text{Model}$ is a summation of the above losses and is minimized using gradient decent~\cite{Adam} to train our ensemble of coordinate MLPs end-to-end. 
\begin{equation}
	\mathbf{L}_\text{Model} = \mathbf{L}_\text{S} + \mathbf{L}_\text{A} + \mathbf{L}_\text{D} + \mathbf{L}_\text{Sym} + \mathbf{L}_\text{K}
\end{equation}

\subsection{Model Fitting}
\label{subsec:modelfitting}
While the aforementioned model can recover interesting anatomic properties of the face with only sparse supervision, it is not very useful unless it can be deformed to match user constraints and serve as a shape prior for an actor facial geometry. 

After training our anatomical implicit face model on a collection of $N$ shapes, the coefficients that are required to deform it include a jaw bone transformation $\mathbf{T_b}^* \in \mathbf{R}^9$, coefficients $\mathbf{w}^* \in \mathbf{R}^{N-1}$ that can be used to blend the corrective expression displacements $\mathcal{B}_e \in \mathbf{R}^{(N-1) \times 3}$, and an optional global head transformation $\mathbf{T_g}^* \in \mathbf{R}^9$. Following equation~\eqref{eq:shape}, we can therefore evaluate our anatomical implicit face model as
\begin{equation}
	\mathbf{s}^* = \mathbf{T_g}^*\left(\text{LBS}\left(\mathbf{\widetilde{s}}_0, \mathbf{T_b}^*, \widetilde{k}\right) + \sum_{N-1}\mathbf{w}^*\mathbf{B}_{e}\right)
\end{equation}
where $\mathbf{T_g}^*$, $\mathbf{T_b}^*$ and $\mathbf{w}^*$ are the only unknowns, and the rest can be queried from a pre-trained AIM. We consider two scenarios for model fitting which include i) fitting our model to a sequence of 3D scans \eg from a facial performance, and ii) fitting our model to 2D landmarks detected on a video~\cite{Chandran2023,wood2022dense}.

For both scenarios, inspired by the state-of-the-art findings of Kim \etal~\cite{Youwang2023NeuFace}, we employ neural reparameterized optimization~\cite{Hoyer2019} and solve for the weights of a simple MLP that predicts the unknown parameters instead of directly optimizing for them. Specifically when given a sequence of $J$ frames with 3D/2D constraints, we optimize for $J$ frame codes $\mathbf{z_j} \in \mathbf{R}^f$ which, when fed as input to a simple 4-layer MLP $\mathbf{F}_T$ with GeLU~\cite{Gelu} activations, predicts the head $\mathbf{T_g}^j$ and jaw $\mathbf{T_b}^j$ poses for each frame. Additionally as the coefficients $\mathbf{w}^j$ are local and spatially varying depending on the template query point $\mathbf{c}$, we use a separate 4-layer MLP $\mathbf{F}_W$ which predicts the coefficients $\mathbf{w}^j$ by taking both the frame code $\mathbf{z_j}$ and the query point $\mathbf{c}$ as input. 
\begin{eqnarray}
	[\mathbf{T_g}^j, \mathbf{T_b}^j] = \mathbf{F}_T(\mathbf{z_j}) \\
	\mathbf{w}^j = \mathbf{F}_W(\mathbf{z_j}, \mathbf{c})
\end{eqnarray}
Unlike the method of Kim \etal~\cite{Youwang2023NeuFace} where the reparameterized optimization was used mainly for improved performance, this neural optimization is even necessary in our case to restrict the number of optimized variables as the number of spatially varying coefficients $\mathbf{w}^*$ used to evaluate our anatomical implicit face model can vary drastically depending on the number of constraint points (see \secref{sec:results}).  

\subsubsection{Fitting Objectives}
\textbf{3D Position Constraint} For fitting our trained model to 3D constraints coming from a facial performance of an actor, we minimize the euclidean distance between the estimated skin point $\mathbf{s}^*$ and the ground truth skin point $\mathbf{s}^{GT}$.
\begin{equation}
	\mathbf{L}_\text{Pos}^{3D} = \lambda_{3D} || \mathbf{s}^* - \mathbf{s}^{GT} ||_{2}^{2}
\end{equation}
\textbf{2D Position Constraint}
For fitting our model to 2D constraints such as facial landmarks estimated by a pre-trained landmark detector \cite{Chandran2023,wood2022dense}, we project the estimated skin point $\mathbf{s}^*$ to screen space using known camera intrinsics $\psi$ and calculate the euclidean distance in 2D between the project point $\psi(\mathbf{s}^*)$ and the corresponding landmark.  
\begin{equation}
	\mathbf{L}_\text{Pos}^{2D} = \lambda_{2D} || \psi(\mathbf{s}^*) - \mathbf{p} ||_{2}^{2}
\end{equation}
$\mathbf{p} \in \mathbf{R}^2$ is a detected landmark corresponding to point $\mathbf{s}^*$.\\[5pt]
\textbf{Coefficient Regularizer}
As the complexity of our implicit anatomical face model can be arbitrarily large, we regularize the estimated blending coefficients $\mathbf{w}^*$ to be small with a weak L2 regularizer.
\begin{equation}
	\mathbf{L}_{W} = \lambda_{Reg}^{w} || \mathbf{w}^* ||_{2}^{2}
\end{equation}
We set $\lambda_{Reg}^{w} = 0.75$ for all our experiments.\\[5pt]
\textbf{Temporal Regularizer}
Finally when optimizing for coefficients on sequential data, we regularize the optimized frame codes $\mathbf{z_{j}}$ to remain similar between adjacent frames. 
\begin{equation}
	\mathbf{L}_{T} = \lambda_{Reg}^{t} || \mathbf{z_j} - \mathbf{z_{j-1}} ||_{2}^{2}
\end{equation}
We set $\lambda_{Reg}^{t} = 0.05$ for all our experiments.\\

Our final fitting energy $\mathbf{L}_\text{Fitting}$ is therefore  
\begin{equation}
	\label{eq:fittingloss}
	\mathbf{L}_\text{Fitting} = \mathbf{L}_\text{Pos}^{3D} + \mathbf{L}_\text{Pos}^{2D} + \mathbf{L}_\text{W} + \mathbf{L}_\text{T} 
\end{equation}
Depending on the 3D or 2D fitting scenario, we set $\lambda_{2D}$ or $\lambda_{3D}$ to 0 respectively. 

\subsection{Implementation Details}
\label{subsec:implementation}
In the model learning stage, we optimize our implicit coordinate networks for $1\mathrm{e}{4}$ iterations with a learning rate of $2\mathrm{e}{-3}$. This takes approximately 10 minutes to converge on a single Nvidia RTX 3090 for an actor model with 40,000 vertices and 20 blendshapes. 

In the model fitting stage, we use a learning rate of $1\mathrm{e}{-3}$ and optimize the fitting MLPs $\mathbf{F}_T$ and $\mathbf{F}_W$ for $1\mathrm{e}{4}$ iterations. This process takes 1 second per frame on a single Nvidia RTX 3090. 

We implement all our MLPs in PyTorch~\cite{pytorch2019}. In our supplementary material we discuss the performance implications of replacing our current python backend with the well engineered fused MLP implementation~\cite{tiny-cuda-nn}.

%% file: 4_results.tex
\section{Results}
\label{sec:results}
We now present several results, applications and evaluations of our \emph{Anatomical Implicit face Model} (AIM).

\begin{figure}[ht]
	\begin{centering}
		\includegraphics[width=\columnwidth]{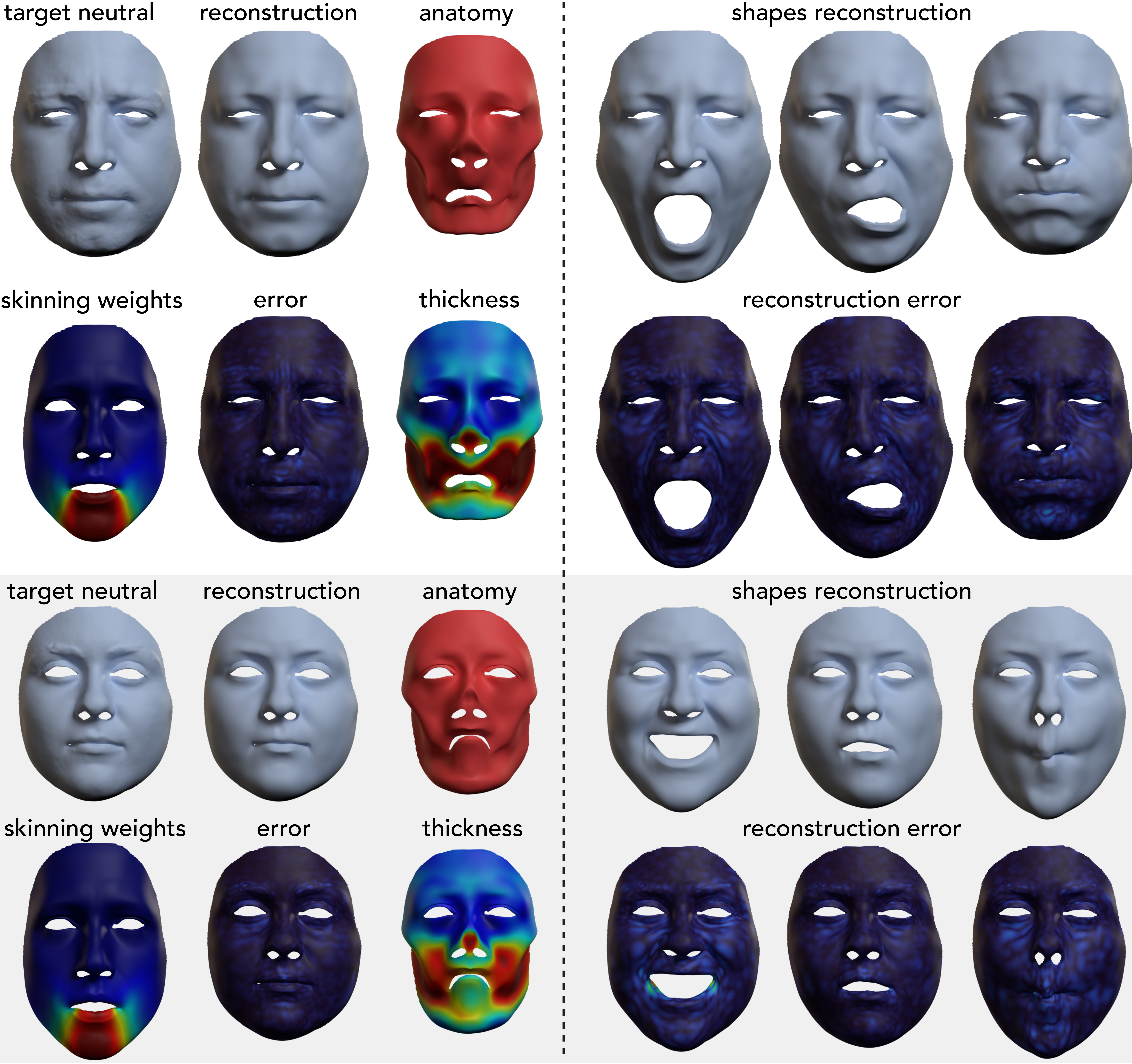}
		\caption{We demonstrate the ability of our \emph{Anatomical Implicit face Model} to recover plausible anatomic features of the face, while also modeling the skin surface with very high fidelity. A subset of 3 expressions from 2 different actor specific models are shown here. The errors are displayed with a scale of 0mm\nobreak \ \includegraphics[width=1.5cm]{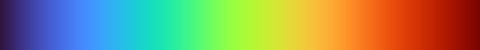}\ 5mm.}
		\label{fig:hero}
	\end{centering}
	\vspace{-2mm}
\end{figure}

\subsection{Learning Actor Specific Anatomy}
We begin by showing the reconstruction accuracy of our AIM on facial blendshapes of multiple actors. As seen in \figref{fig:hero} on 2 different actors, our method can faithfully represent facial shapes with high fidelity while capturing both the low and high frequency features of facial shape and expression. We also show the anatomic features recovered by our new formulation which includes the dense underlying facial anatomy (shown in red), the soft tissue thickness at every point on the anatomy  (visualized as heatmap), and the optimized subject specific skinning weights. These results highlight the new abilities introduced by our method in recovering plausible anatomy features while jointly learning to model surface deformations.

\subsection{Anatomy Manipulation}
\begin{figure}
	\begin{centering}
		\includegraphics[width=0.9\columnwidth]{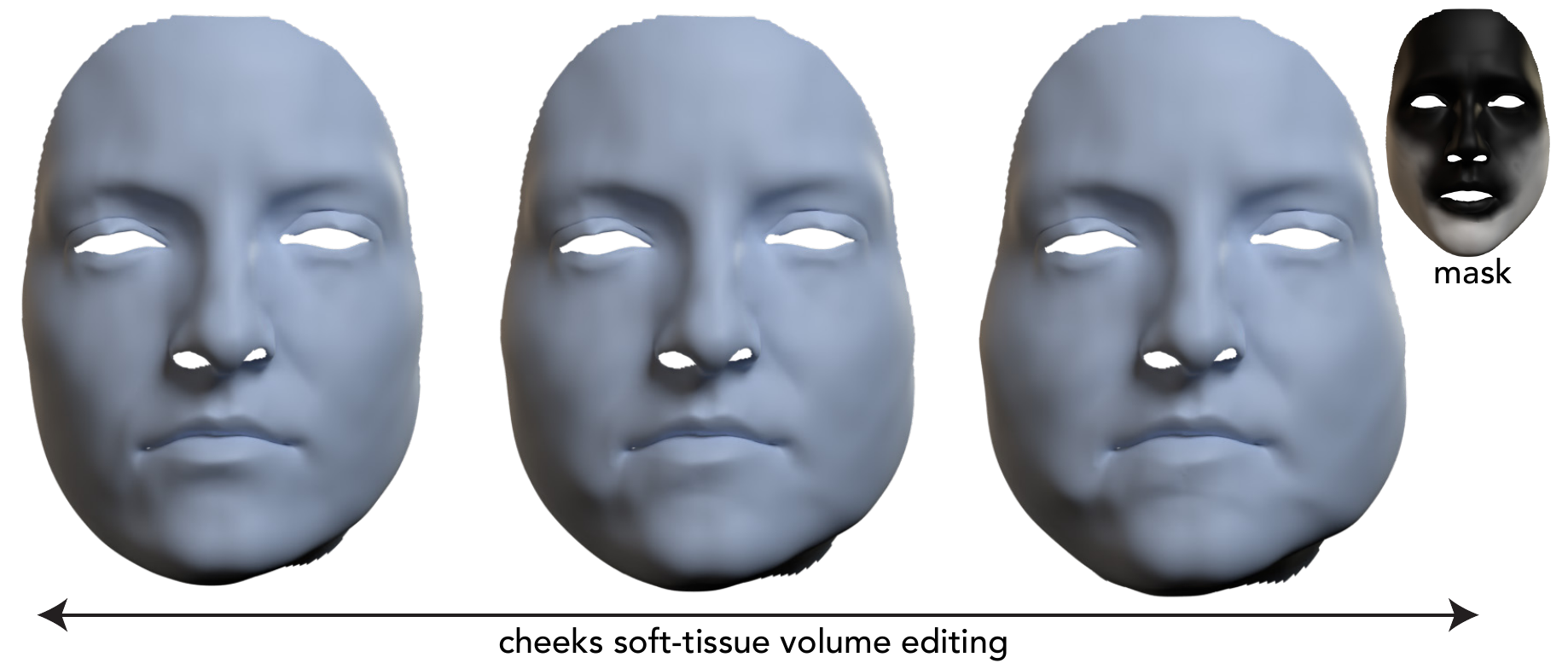}
		\caption{Once the AIM is learned for an actor, it can be used to intuitively deform a face using the learned anatomic properties, as demonstrated here by scaling the soft tissue thickness in a hand painted cheek region, and by propagating the change to the skin surface thanks to our formulation.}
		\label{fig:anatomyManip}
	\end{centering}
	\vspace{-5mm}
\end{figure}

Our ability to estimate the underlying anatomy that densely constrains the skin surface opens up new, yet computationally inexpensive ways to edit facial geometry using our learned anatomic properties. For example, as illustrated in \figref{fig:anatomyManip}, by simply scaling the learned soft tissue thickness $d$ in desired regions of the face (denoted by the hand drawn mask), an artist can interactively sculpt/deform an actor's face shape to match their requirements.

\subsection{Expression Reconstruction}
\begin{figure}
	\begin{centering}
		\includegraphics[width=\columnwidth]{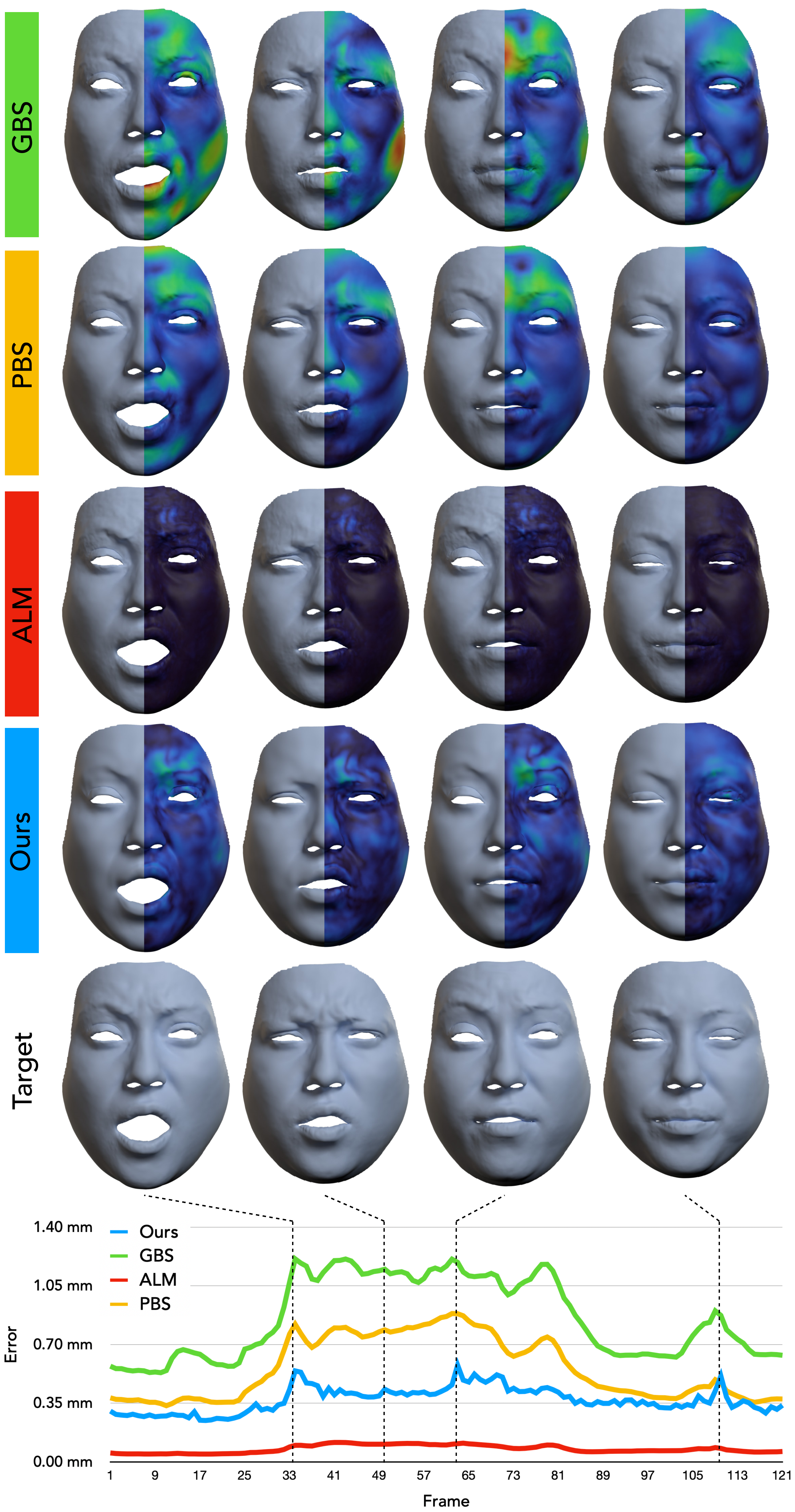}
		\caption{We show qualitative and quantitative comparisons of fitting 3D performances with various actor specific models. All the errors are displayed with a scale of 0mm\ \includegraphics[width=2cm]{./colormap.png}\ 5mm.}
		\label{fig:3dfitting}
	\end{centering}
	  \vspace{-3mm}
\end{figure}

\begin{figure}
	\begin{centering}
		\includegraphics[width=\columnwidth]{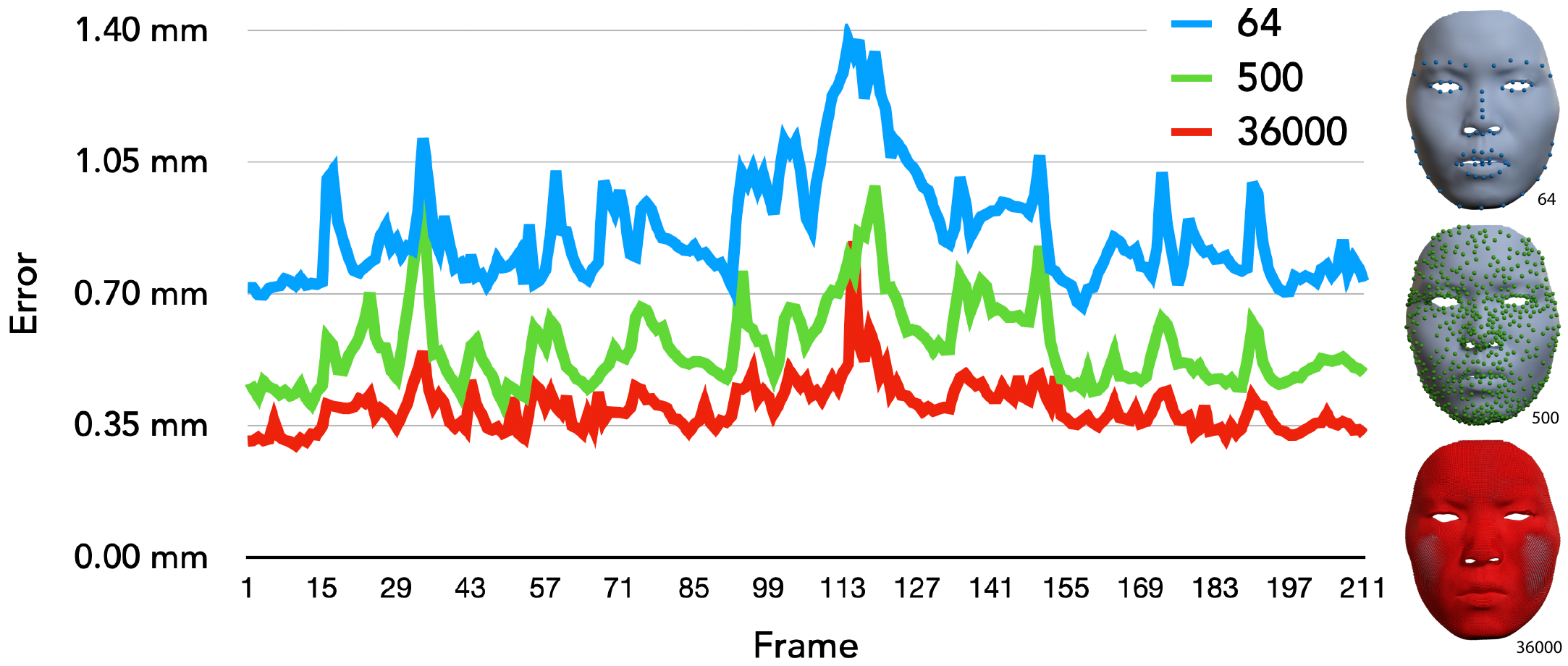}
		\caption{Our continuous anatomical face model can be fit to 3D scans with varying density of constraints and still provide valid results due to our fitting algorithm.: all the errors are displayed with a scale of 0mm \includegraphics[width=2cm]{./colormap.png} 5mm.}
		\label{fig:sparsity}
	\end{centering}
\end{figure}

We next evaluate the expressiveness of our model by fitting it to unseen 3D performances of multiple actors. Given a sequence of $J$ dynamic 3D shapes from a studio scanner~\cite{Beeler:2011}, we first deform our template mesh $C$ to match the scanned shapes using standard mesh registration techniques such that the dynamic 3D scans are in full vertex correspondence with our AIM. We then follow the fitting procedure described in \secref{subsec:modelfitting} and obtain per-frame transformations $[\mathbf{T_g}^j, \mathbf{T_b}^j]$ and shape coefficients $\mathbf{w}^j$ that explain the captured ground truth shape. For this experiment, we use the 3D position constraint from Eq.~\eqref{eq:fittingloss} and set $\mathbf{L_{2D}}$ to 0. We densely constrain the fitting procedure at every vertex of the ground truth shape. In \figref{fig:3dfitting} we provide both a qualitative and quantitative comparison of fitting to novel performance from an actor against global blendshapes (GBS)~\cite{Lewis2014}, a patch blendshape model (PBS)~\cite{Chandran2023}, and the anatomical local model (ALM)~\cite{Wu2016ALM}. In this experiment, we use 20 ground truth actor blendshapes to build the GBS, PBS, and ALM models, and the anatomically reconstructed blendshapes for our method. Even under this slight disadvantage, our method outperforms both GBS, and PBS and provides visually comparable results to the ALM model. \tabref{tab:errors} shows the average fitting error of each method across 819 frames from 5 sequences of 5 different actors. Our method converges in a few seconds for each frame, while the ALM algorithm consistently requires several minutes per frame. While the continuous nature of AIM enables us to evaluate it with coefficients of arbitrary locality, it could result in situations where our fitting is underconstrained in the absence of dense constraints leading to broken shapes. To illustrate that this does not happen in our reparameterized optimization, we show the result of fitting the AIM to sparse constraints in \figref{fig:sparsity}. While increasing the density of constraints improves the fitting accuracy, fitting our model to sparse landmarks also provides plausible results. Note that we do not compare fitting accuracy against generic morphable models like FLAME~\cite{FLAME2017} or NPHM~\cite{giebenhain2023nphm} as ours is actor specific and therefore a quantitative comparison might be unfair to the other methods. However we present some qualitative comparison to generic models in our supplemental material.

\begin{table}
  \centering
  \caption{Average fitting error across 819 frames from 5 sequences of 5 different actors. See supplemental material for details.}
  \label{tab:errors}
  \begin{tabular}{|c|c|c|c|}
    \hline
    \textbf{GBS}~\cite{Lewis2014} & \textbf{PBS}~\cite{Chandran2023} & \textbf{ALM}~\cite{Wu2016ALM} & \textbf{Ours} \\
    \hline
    0.834 mm & 0.51 mm & 0.095 mm & 0.312 mm \\
    \hline
  \end{tabular}
  \vspace{-5mm}
\end{table}

\subsection{3D Performance Retargeting}
\begin{figure}
	\begin{centering}
		\includegraphics[width=\columnwidth]{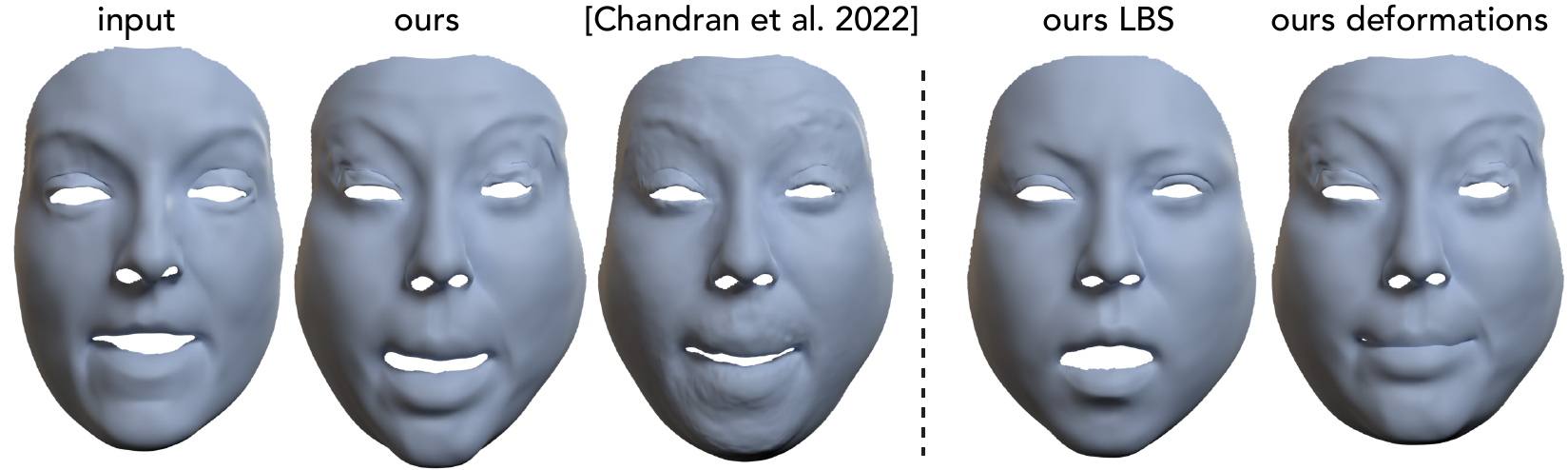}
		\caption{We show the result of facial performance transfer in 3D from an input actor (left) to a different actor as produced by our method (2nd column) and the local retargeting model of Chandran \etal \cite{Karacast2022}. While providing qualitatively similar results, our model implicitly disentangles the performance into rigid jaw motion (3rd column), and nonrigid soft tissue deformations (4th column).}
		\label{fig:retargeting}
	\end{centering}
\end{figure}

Another important application of our method is in the area of 3D performance retargeting, where the goal is to transfer a facial animation from a source to a target character while respecting the identity and anatomic characteristics of the target character. To accomplish this using our model, we learn two separate instances of our model for the source and target character respectively from a sparse set of 20 blendshapes in correspondence. We then fit our source model to the facial animation of the source target character to obtain per-frame transformations $[\mathbf{T_g}^j, \mathbf{T_b}^j]$ and shape coefficients $\mathbf{w}^j$. These coefficients can simply be played back on the target model to achieve facial performance retargeting. In \figref{fig:retargeting}, we provide a qualitative comparison to the state-of-the-art 3D retargeting algorithm of Chandran \etal~\cite{Karacast2022} by retargeting the performance from a source to a target character. Our method provides competitive results to state of the art, while also allowing users to disentangle the rigid jaw motion and the soft tissue deformations of the skin surface. Our method additionally provides a substantial runtime benefit here and retargets each frame in a few (2-3) seconds, while the method of Chandran \etal requires several minutes per frame due to a costly anatomic solve. Finally unlike the approach of Chandran \etal, our method provides all of above benefits without having to manually choose design parameters such as the patch layout, number of overlaps etc.

\subsection{Ablations}
\label{subsec:ablation}

\begin{table}
  \centering
  \caption{Average error in mm on a sequence of 100 frames using different types of activation functions in our MLPs.}
  \label{tab:ablationActivation}
  \begin{tabular}{|c|c|c|c|}
    \hline
    \textbf{gelu} & \textbf{relu} & \textbf{siren} \\
    \hline
    0.71 mm & 0.62 mm & 0.21 mm \\
    \hline
  \end{tabular}
    \vspace{-5mm}
\end{table}

\begin{table}
  \centering
  \caption{Average error in mm on a sequence of 80 frames using variation of our loss functions during the model learning stage.}
  \label{tab:ablationLosses}
  \begin{tabular}{|c|c|c|c|c|c|}
    \hline
    \textbf{no $\mathbf{L}_\text{A}$} & \textbf{no $\mathbf{L}_\text{K}$} & \textbf{no $\mathbf{L}_\text{Sym}$} & \textbf{no $\mathbf{L}_\text{D}$} & \textbf{$\mathbf{L}_\text{Model}$ (Ours)} \\
    \hline
    0.29 & 0.22 & 0.24 & 0.21 & 0.19 \\
    \hline
  \end{tabular}
    \vspace{-5mm}
\end{table}

Finally \tabref{tab:ablationActivation} shows an ablation study on our choice of activation in our MLPs and \tabref{tab:ablationLosses} shows an ablation on the effect of several of our loss functions on the recovered geometry. Additional ablations are provided in the supplemental material.

%% file: 5_conclusion.tex
\section{Conclusion}
\label{sec:conclusion}
In this paper we propose a new anatomically constrained implicit face model which provides a holistic representation of both facial anatomy and the enclosing skin surface using an ensemble of coordinate neural networks. Given an arbtrary set of skin surface meshes and only a neutral shape with estimated skull and jaw bones, our method recovers a dense anatomical substructure to constrain each point on the skin surface, and can model complex skin deformations with high fidelity. While we have explored the use of such a model in the context of actor specific blendshape models, future work could analyze it's implications as a generic morphable model, by extending our formulation to handle multiple identities at once. Our new \emph{Anatomical Implicit face Model} (AIM) has applications in shape representation and manipulation, retargeting and more, and we hope that our method encourages exciting future research.  

%% file: supplemental.tex
\clearpage
\setcounter{page}{1}
\maketitlesupplementary

\section{Additional Details}
\subsection{Anatomy Constraints}
\begin{figure}
	\begin{centering}
		\includegraphics[width=\columnwidth]{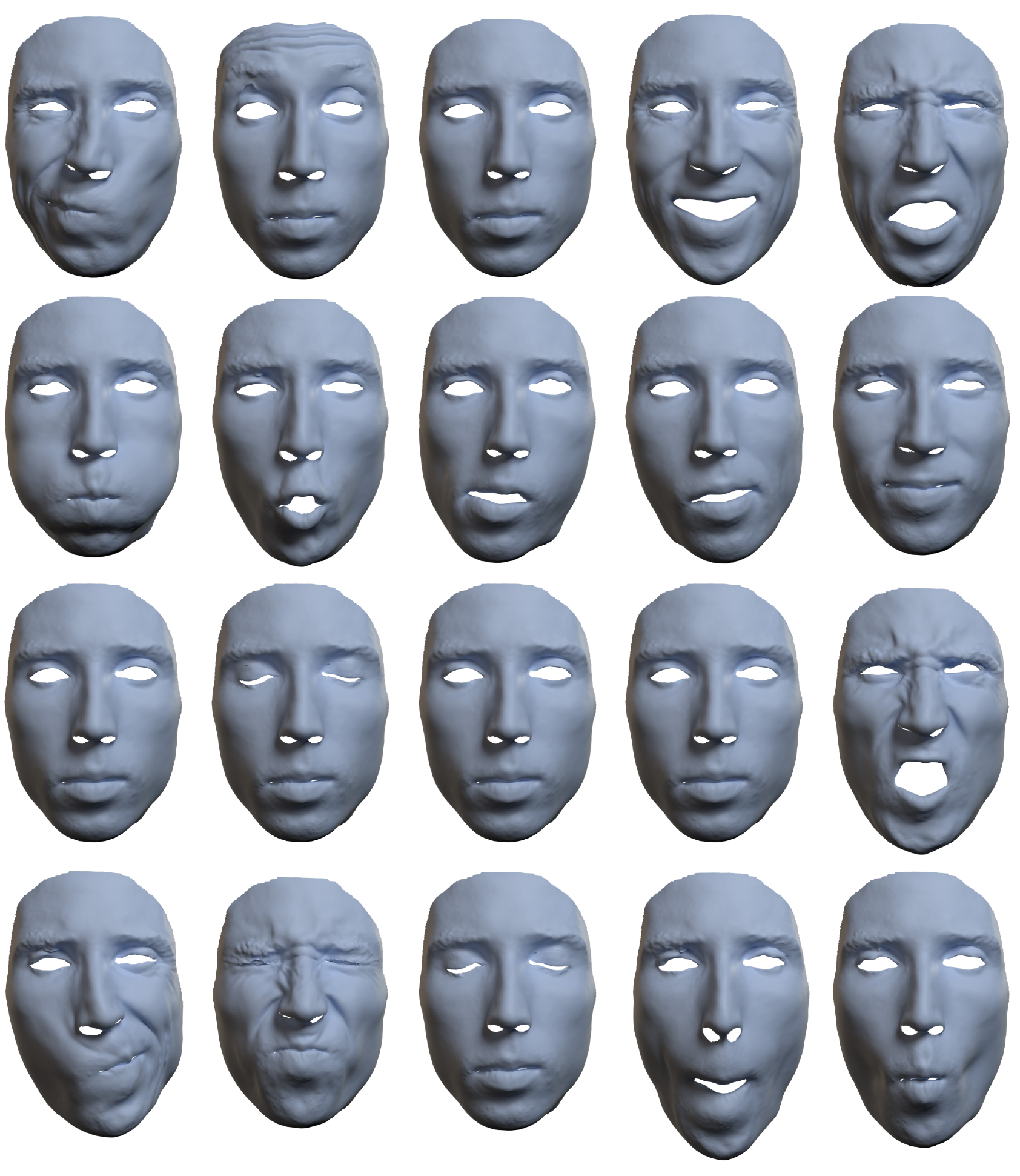}
		\caption{We show here the collection of 3D shapes used in our Model Learning stage. For all our experiments we used 1 neutral expression (or rest pose) and 19 expressions, all captured and reconstructed following the method of Beeler \etal~\cite{Beeler:2011}.}
		\label{fig:allExpressions}
	\end{centering}
\end{figure}

\begin{figure}
	\begin{centering}
		\includegraphics[width=\columnwidth]{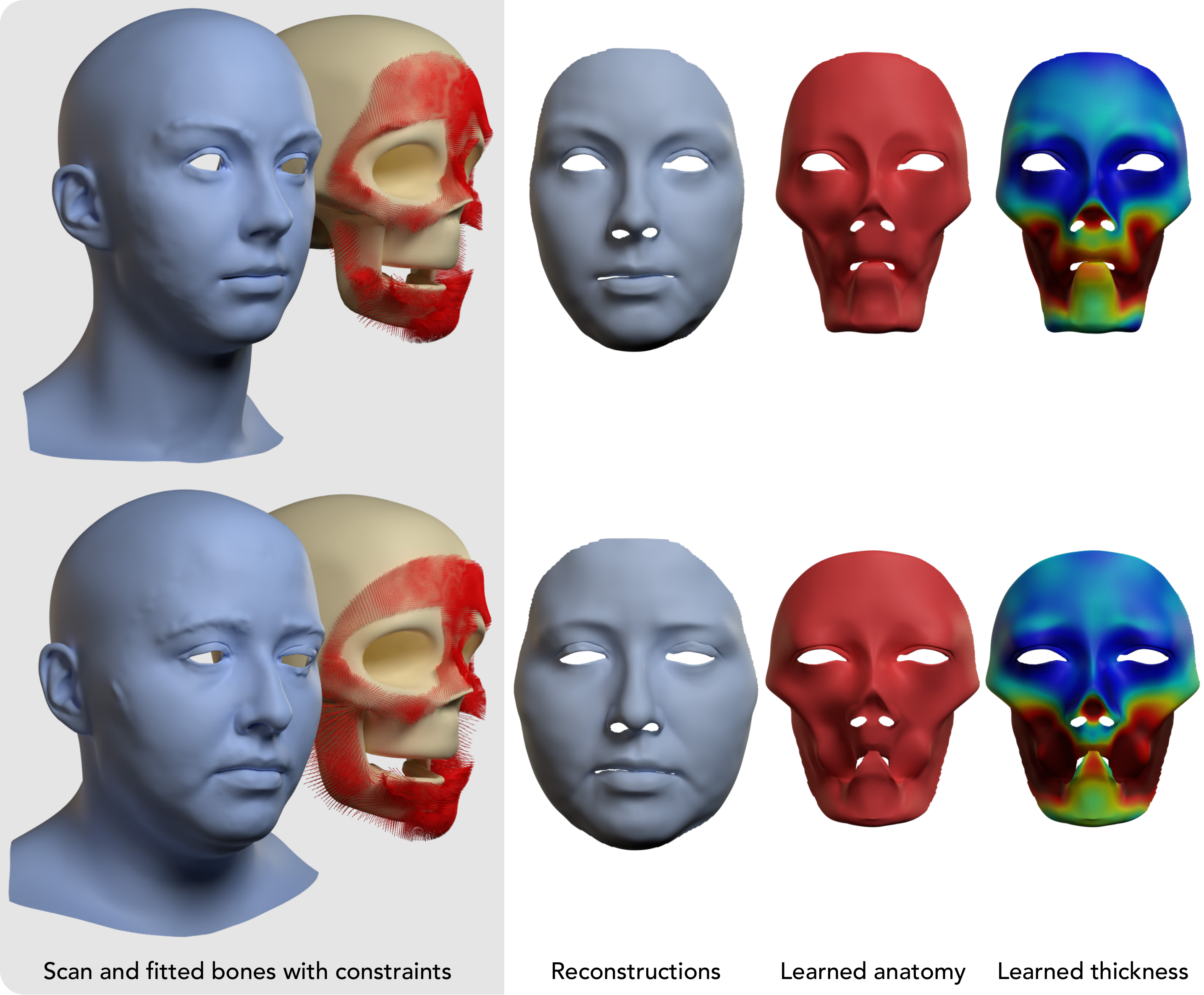}
		\caption{We show for two actors, first on the left the input neutral geometry next to the fitted skull and mandible, with an overlay of our computed sparse anatomical constraints. On the right, we show the reconstructed geometry, the learned anatomy (using those sparse anatomical constraints) and learned thicknesses.}
		\label{fig:anatomyConstraints}
	\end{centering}
\end{figure}

We loosely regularize the skull and mandible geometries using sparse anatomical constraints. We compute these sparse constraints by fitting a template skull and mandible meshes to the neutral geometry following the method of Zoss \etal~\cite{Zoss2018}. For any given skin point inside a hand-painted trusted region of the bone fitting process, we trace a ray along the inverse direction of the skin normal and store the bone intersection point only if the bone faces the same direction as the skin. We then trace another ray following now the bone normal, intersecting the skin again (potentially at a different point) and store the thickness and bone normal for the intersected skin point. Overall our sparse anatomical constraints exist only for 5 to 10\% of the skin query points. We then use those bone points and thicknesses inside our losses $\mathbf{L}_\text{A}$ and $\mathbf{L}_\text{D}$ respectively. We show a visualization of the anatomical constraints and learned anatomies and thicknesses on~\figref{fig:anatomyConstraints}. A visual depiction of the full set of 20 shapes used in our work is shown in \figref{fig:allExpressions}.

\subsection{Network Architecture}

\begin{figure*}
	\begin{centering}
		\includegraphics[width=\textwidth]{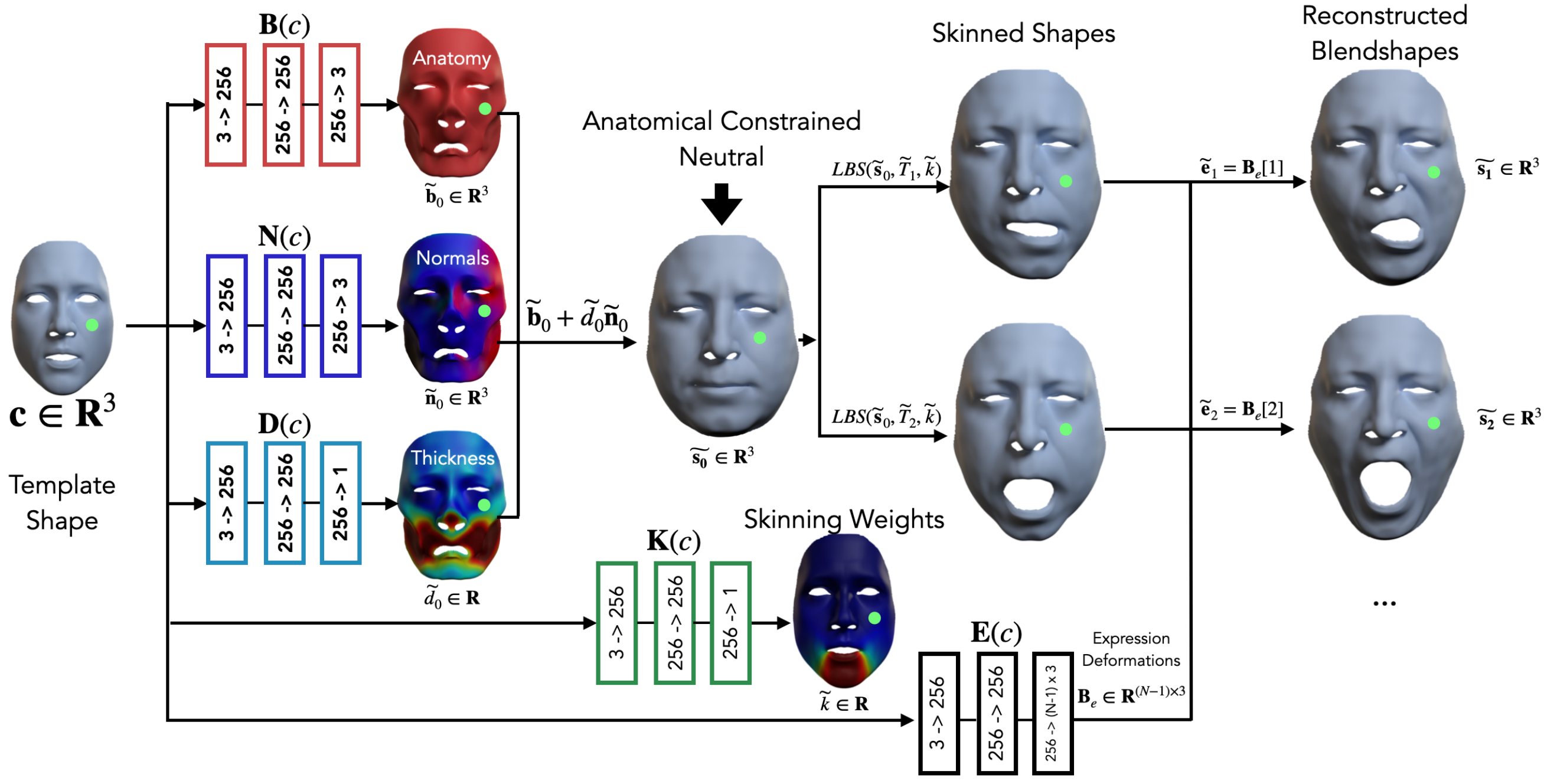}
		\caption{Starting from a query point $\mathbf{c}$ on the template shape, an ensemble of Siren MLPs \cite{sitzmann2019siren} predict the dense underlying anatomy $\widetilde{\mathbf{b}}_0$, anatomy normals $\widetilde{\mathbf{n}}_0$, and the soft tissue thickness $\widetilde{d}_0$, using which a neutral shape $\widetilde{\mathbf{s}}_0$ of the actor is reconstructed. Then using learned per-shape jaw transformations $\widetilde{T}_i$, and actor specific skinning weights $\widetilde{k}$, the neutral is skinned to account for the rigid jaw movement. Finally, expression specific deformations $\widetilde{\mathbf{e}}_i$ are added on top of the skinned mesh to reconstruct the given blendshapes.}
		\label{fig:memorizationNetwork}
	\end{centering}
\end{figure*}

\begin{figure*}
	\begin{centering}
		\includegraphics[width=\textwidth]{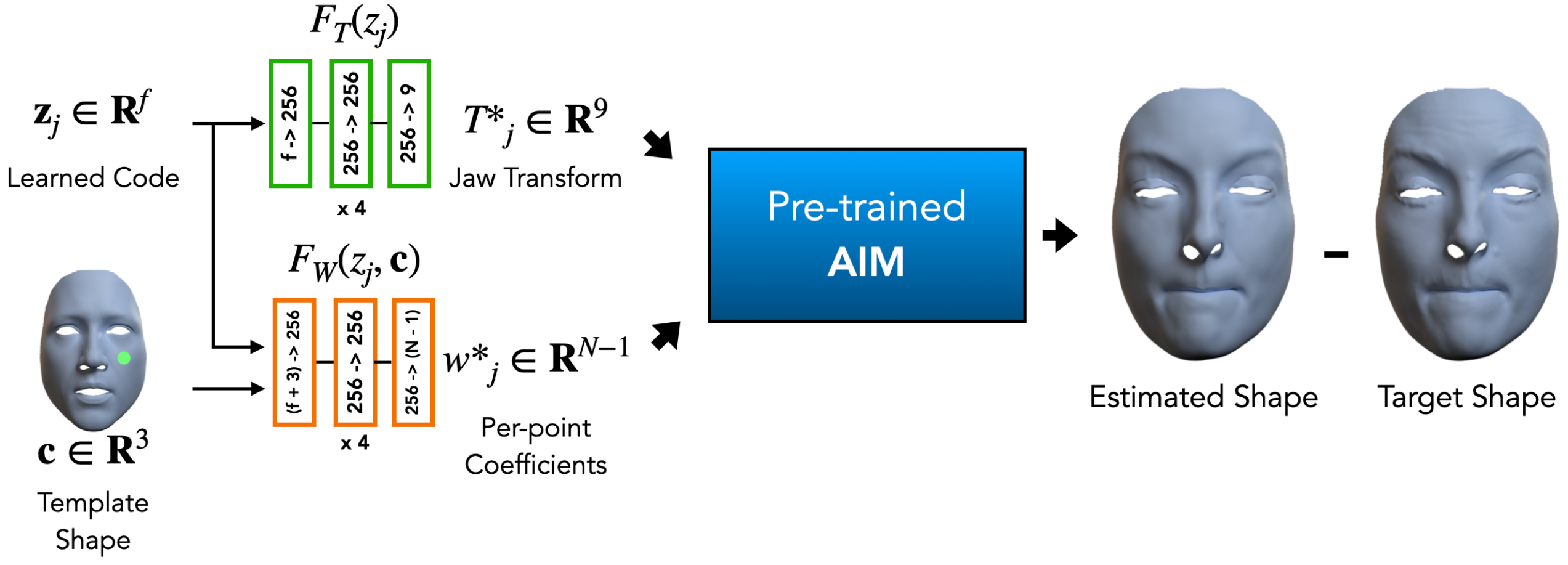}
		\caption{Given a query point $\mathbf{c}$ and a learned code $\mathbf{z}_j$  for each target shape, we use small fitting MLPs to predicts the jaw transformation $\widetilde{T}_j^*$ and the per-point coefficients $\widetilde{\mathbf{w}}_j^*$, using which the AIM model can be evaluated to result in the estimated shape. The fitting MLPs are trained to minimize the reconstruction error between the estimated and target shape. }
		\label{fig:fittingNetwork}
	\end{centering}
\end{figure*}

In \figref{fig:memorizationNetwork} and \figref{fig:fittingNetwork}, we show a detailed breakdown of our memorization and fitting networks.   

\section{Additional Results}
\subsection{Face Reconstruction from 2D Landmarks}
\begin{figure}[ht]
	\begin{centering}
		\includegraphics[width=\columnwidth]{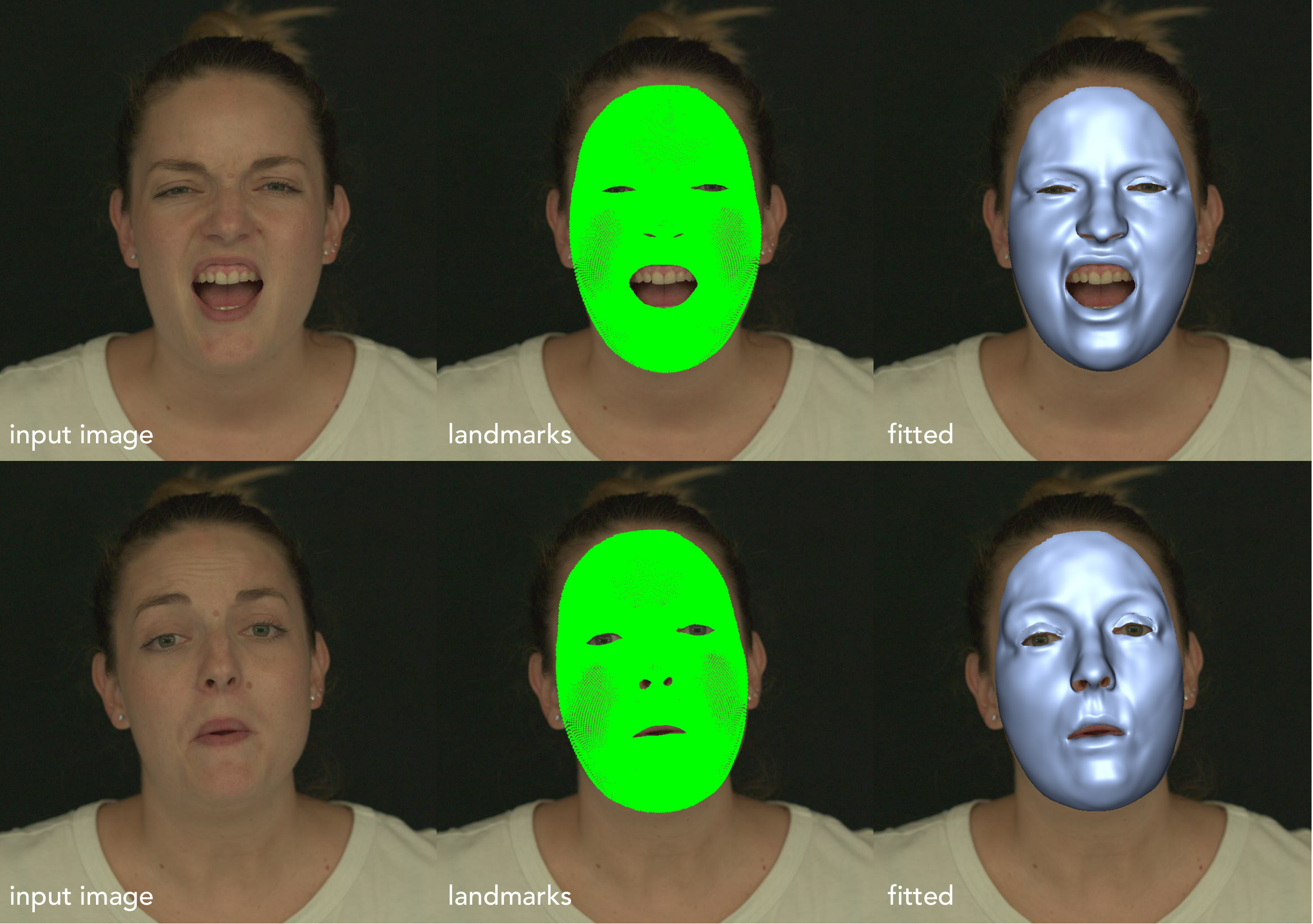}
		\caption{We demonstrate a proof of concept of the application of our model in face reconstruction, where our AIM model can be fit to 2D landmarks obtained from a pre-trained landmark detector, capturing both the pose and expression of the person faithfully. }
		\label{fig:2dfitting}
	\end{centering}
\end{figure}

In the main paper, we describe how to formulate a 2D position constraint to fit our anatomical implicit face model to landmarks obtained from a pre-trained landmark detector. In \figref{fig:2dfitting}, we show qualitative results of fitting our trained anatomical implicit model to 10,000 dense landmarks predicted by a 2D landmark detector \cite{Chandran2023} on an input monocular video.  
\subsection{Learning Actor Specific Anatomical Properties}
\begin{figure*}
	\begin{centering}
		\includegraphics[width=0.8\textwidth]{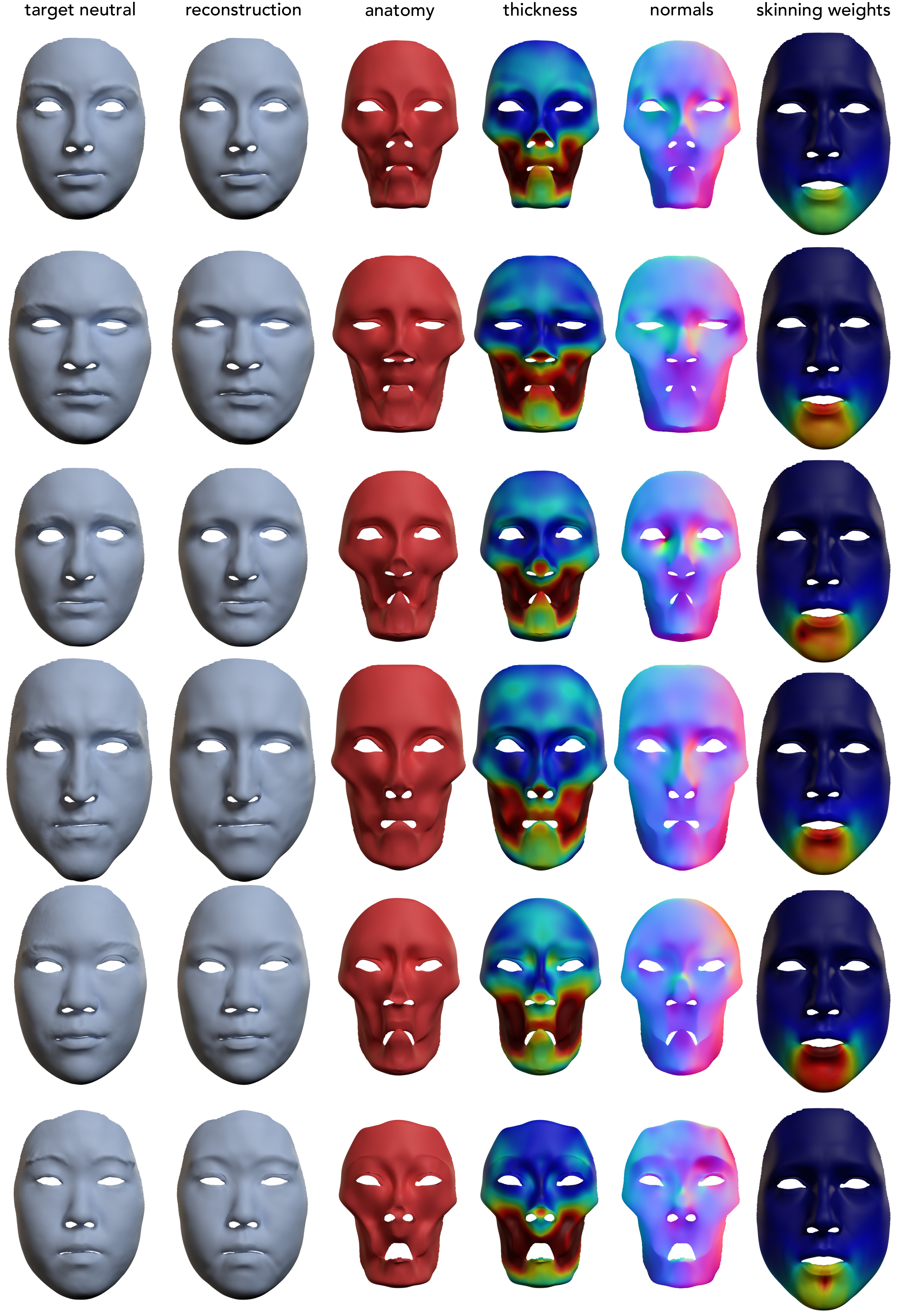}
		\caption{We show the anatomical features recovered by our formulation across a wide variety of actors. From left to right, we show the ground truth neutral shape, the reconstructed neutral shape, our learned anatomy, our learned soft-tissue thickness, our learned anatomical normals, and our learned subject specific skinning weights.}
		\label{fig:anatomicalProperties}
	\end{centering}
\end{figure*}

In \figref{fig:anatomicalProperties}, we show additional results of the recovered dense anatomical properties on a number of actors with varying face shapes spanning different ethnicities, and age groups. 
\subsection{Runtime Analysis}
Our model fitting stage, which involves the training of the fitting MLPs $F_W$ and $F_T$  (see the main text), takes atmost a few seconds per-frame to converge on a Nvidia RTX 3090. As an engineering update to our system, we experimented with the \emph{tinycuda} framework of Muller \etal \cite{tiny-cuda-nn} and found that it provided a 2x performance improvement in model fitting, without any adverse effects on fitting accuracy. We leave a more thorough performance optimization of our pipeline to future work, which could also include exploring fused MLPs for the model learning stage.  

\subsection{3D Performance Retargeting}
We kindly refer you to our supplemental video for additional retargeting results and qualitative comparisons.

\subsection{Ablations}
\begin{figure}
	\begin{centering}
		\includegraphics[width=\columnwidth]{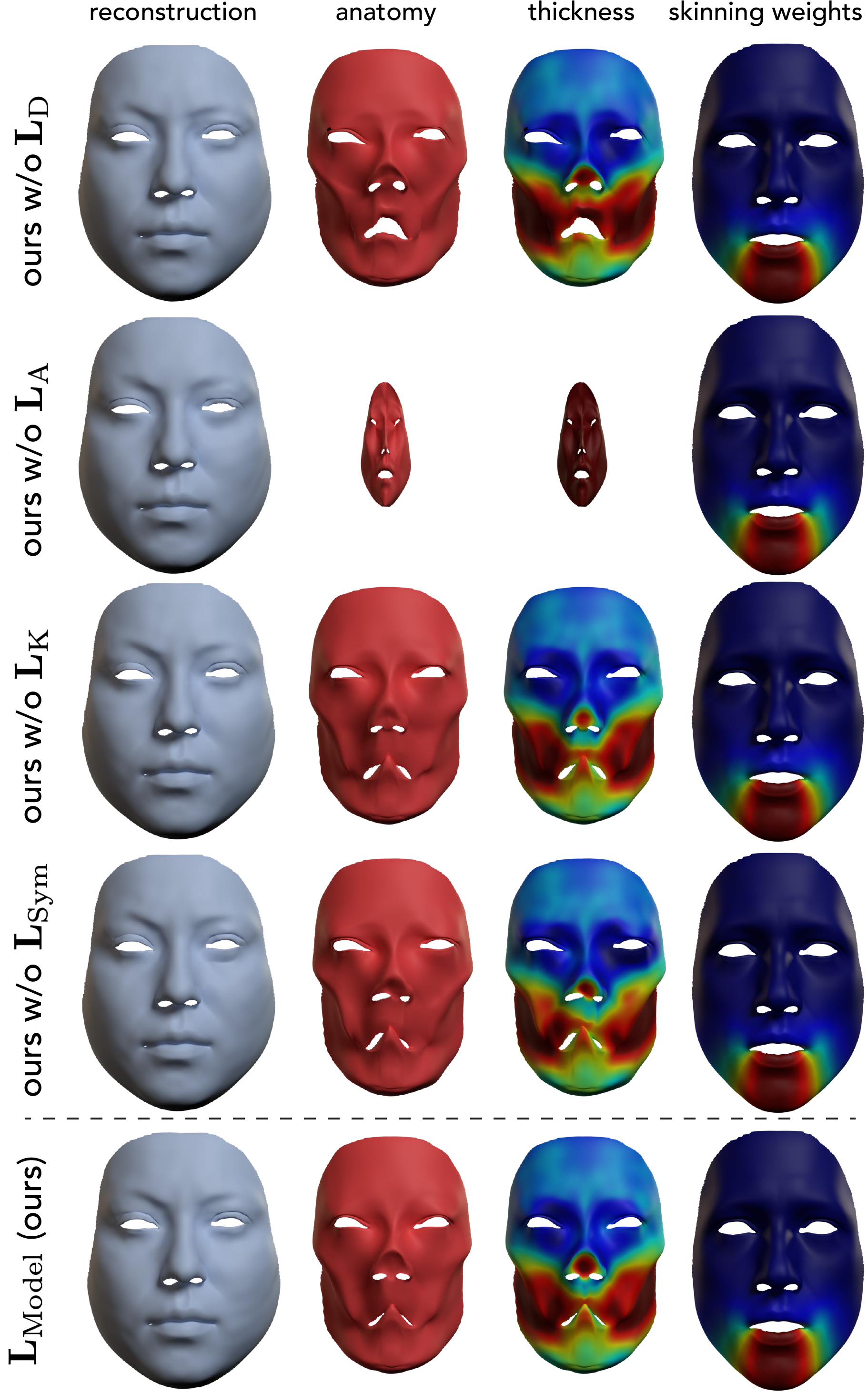}
		\caption{1st row: We show the effect of removing the thickness regularizer $L_D$ that encourages the soft tissue thickness to remain small in unconstrained areas, 2nd row: the effect of removing the anatomy loss $L_A$ which result in a collapse of the learned anatomy, while still reconstructing the neutral in the first column, 3rd row: The effect of removing the optional skinning weight regularizer $L_K$ which does not adversely affect the learned skinning weights as seen in the last column, 4th row: The effect of removing the symmetry regularizer on the anatomy, as a result of which the anatomy no longer remains symmetric, and last row, our $L_{\text{Model}}$ loss which uses a weight sum of all regularizers.} 
		\label{fig:lossAblation}
	\end{centering}
\end{figure}
\begin{figure}
	\begin{centering}
		\includegraphics[width=\columnwidth]{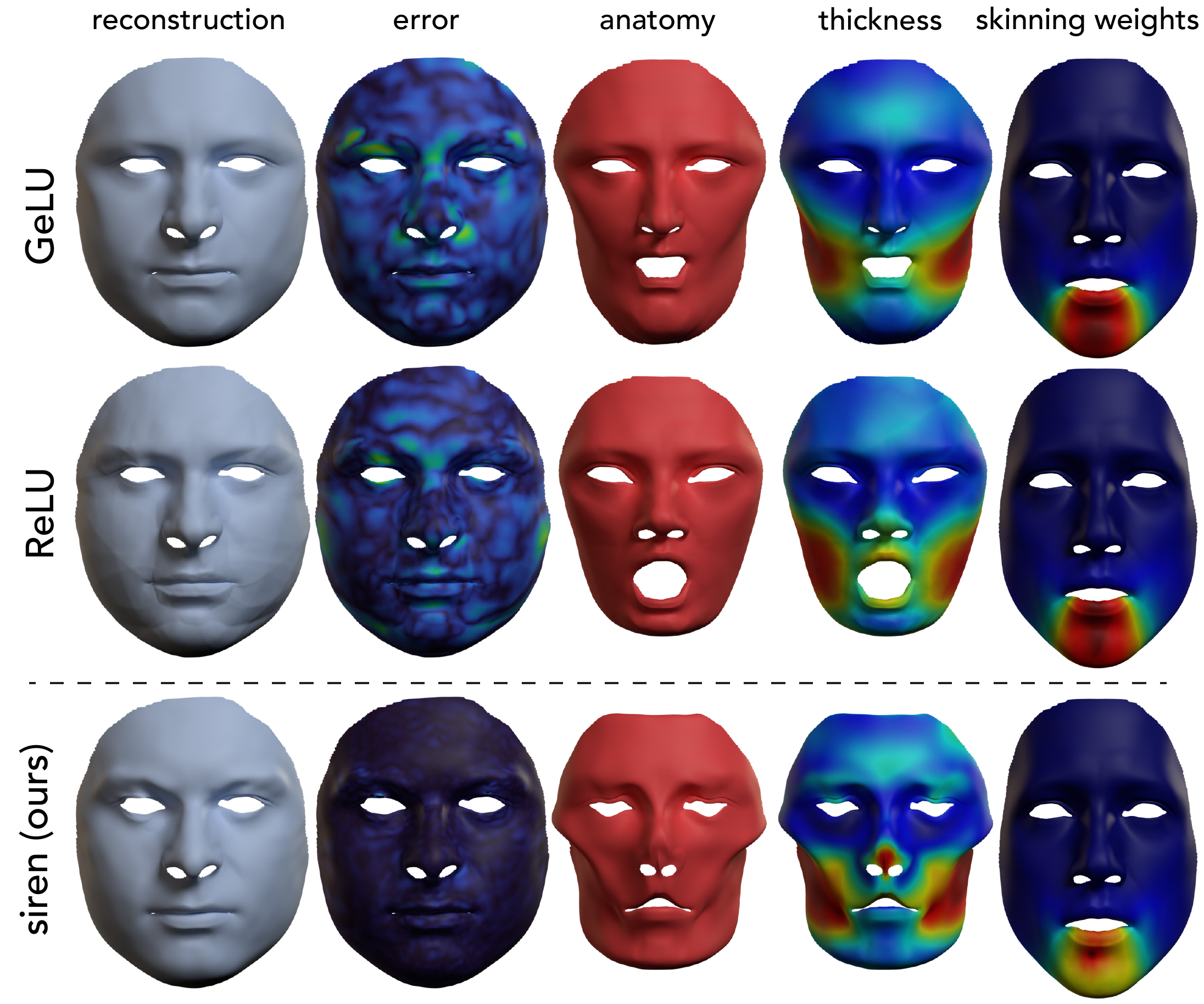}
		\caption{Using GeLU and ReLU activations in our implicit MLPs results in oversmoothed anatomy and reconstructions lacking surface detail. Sine activations provided the best results.}
		\label{fig:activationAblation}
	\end{centering}
\end{figure}
\begin{figure}
	\begin{centering}
		\includegraphics[width=0.95\columnwidth]{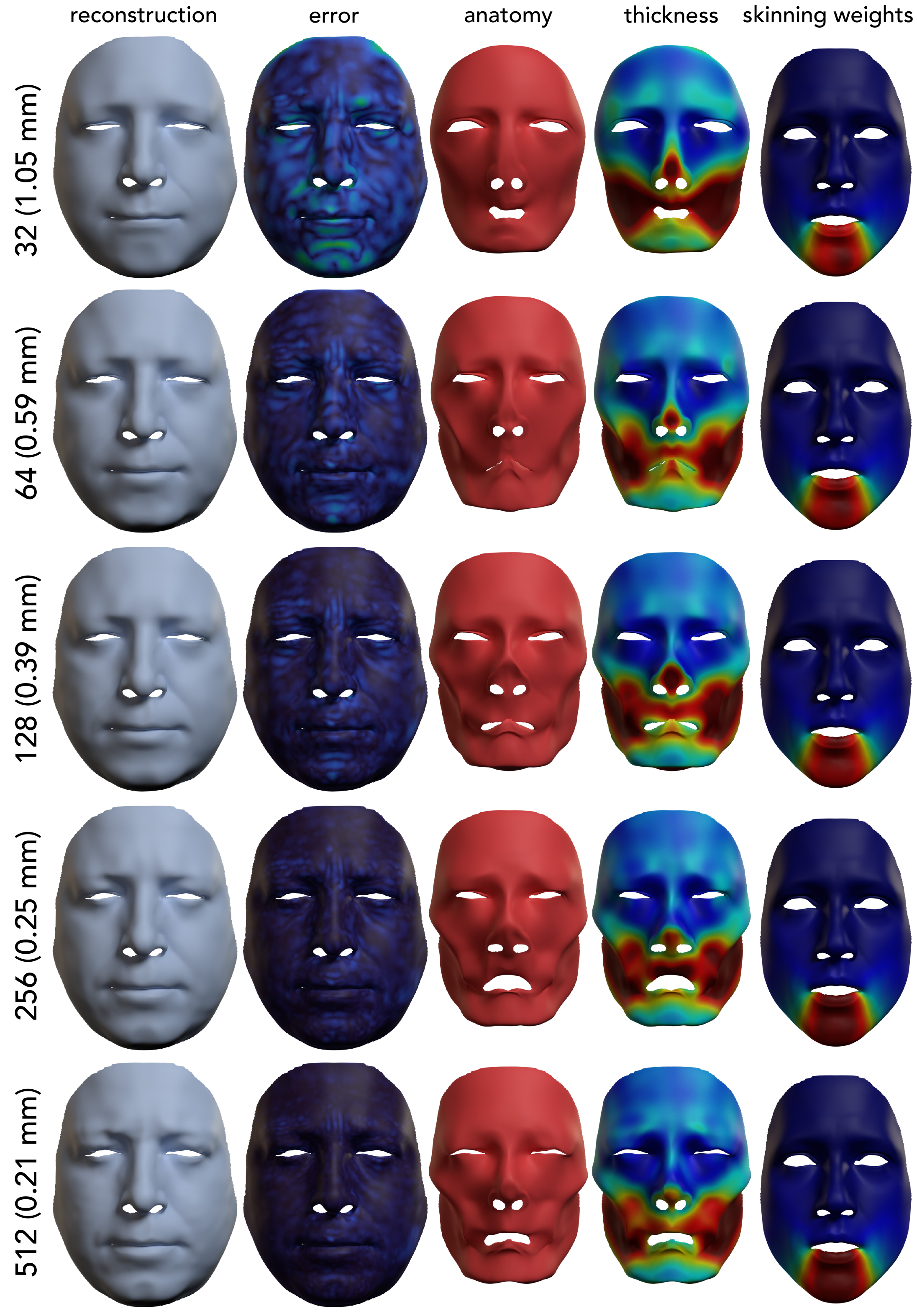}
		\caption{While increasing the size of the hidden layers in our MLPs improved reconstruction performance, it comes at the cost of a larger network that is slower to evaluate. In our work, we used a hidden layer size of 256 neurons which provided a good balance between accuracy and performance.}
		\label{fig:hiddenLayerSizeAblation}
	\end{centering}
\end{figure}

We provide visual results for the several ablations we performed in our work, which include the effect of removing certain regularizers used during the model learning stage (see section in the main text) in~\figref{fig:lossAblation}, the effect of different activation functions in~\figref{fig:activationAblation}, and the size of the hidden layers used during model learning in~\figref{fig:hiddenLayerSizeAblation}. 

\subsection{Generic Model Comparison}
\begin{figure}[ht]
	\begin{centering}
		\includegraphics[width=\columnwidth]{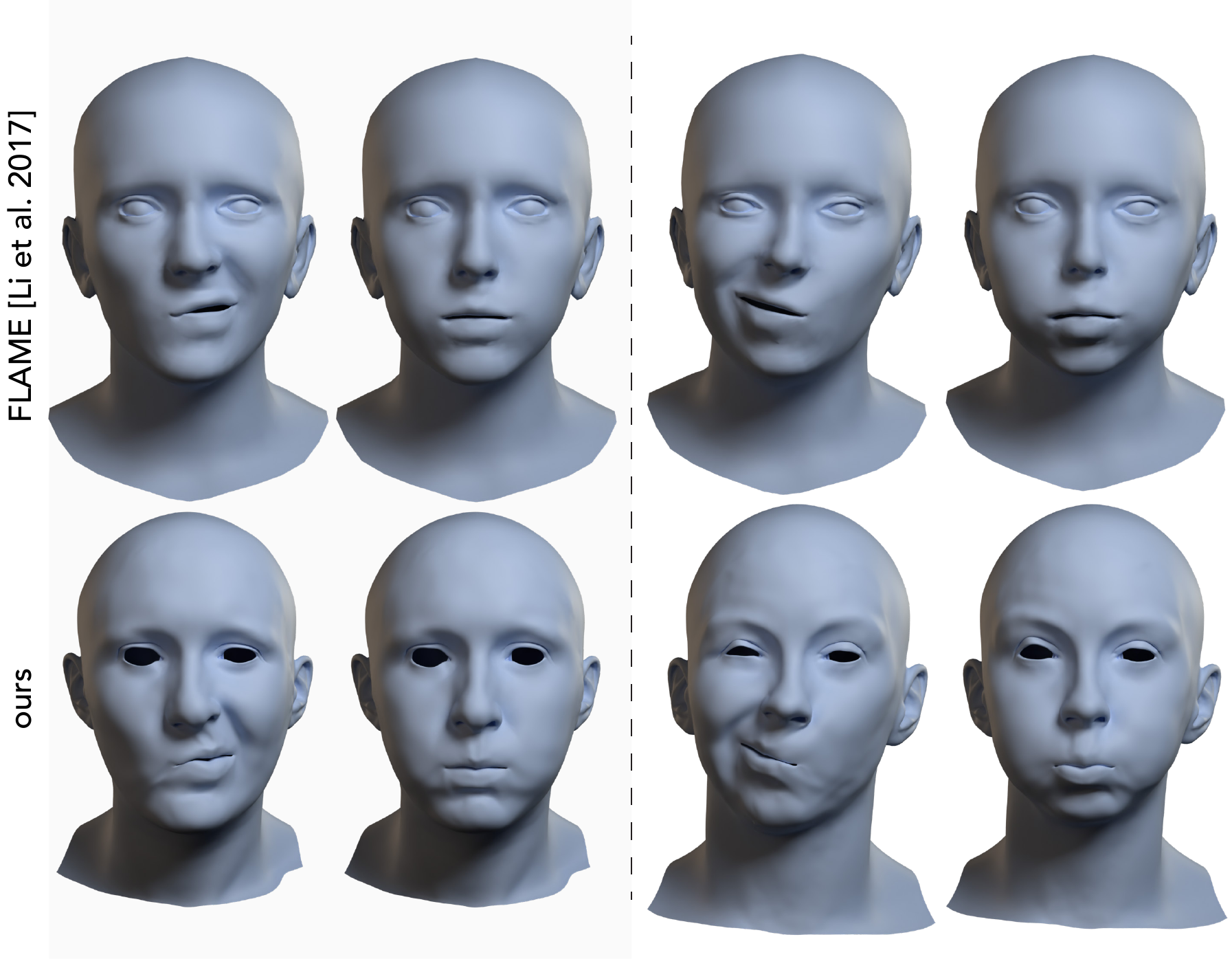}
		\caption{We show 2 expression of 2 different actors fitted by our model and the FLAME model~\cite{FLAME2017}. A generic 3DMM is unable to faithfully capture an particular individuals shape that lies outside of it's shape space.}
		\label{fig:comparisonWithFlame}
	\end{centering}
\end{figure}

As discussed in the main text, a quantitative comparison of our actor specific model against a generic 3D morphable model would be unfair to general 3DMMs as they serve a more diverse purpose. However in~\figref{fig:comparisonWithFlame} we show a visual comparison of 2 expressions fitted using 3D positions as constraints with our model and the FLAME model~\cite{FLAME2017} for 2 different actors. 

